\documentclass[a4paper,11pt,fleqn]{article}
\usepackage{amsmath,amssymb,amsthm} 

\newcommand{\rank}{\mathop{\rm rank}\nolimits}
\newcommand{\Inv}{{\rm Inv}}

\newcommand{\NilRad}{\mathop{\rm NR}\nolimits}
\newcommand{\diag}{\mathop{\rm diag}\nolimits}

\newcommand{\todo}[1][\null]{\ensuremath{\clubsuit}}
\newcommand{\noprint}[1]{}

\newtheorem{theorem}{Theorem}
\newtheorem{lemma}[theorem]{Lemma}
\newtheorem{corollary}[theorem]{Corollary}
\newtheorem{proposition}[theorem]{Proposition}
{\theoremstyle{definition}
\newtheorem{remark}[theorem]{Remark}
}

\begin{document}

\allowdisplaybreaks


\par\noindent {\LARGE\bf
Invariants of solvable Lie algebras with triangular\\ nilradicals and diagonal nilindependent elements
\par}

\vspace{4mm}\par\noindent {\large 
Vyacheslav Boyko~$^\dag$, Jiri Patera~$^\ddag$ and Roman O. Popovych~$^{\dag\S}$
} \par\vspace{2mm}\par

\vspace{2mm}\par\noindent {\it
$^\dag$~Institute of Mathematics of NAS of Ukraine, 3
Tereshchenkivs'ka Str., Kyiv, 01004 Ukraine}\\
$\phantom{^\dag}$~E-mail: boyko@imath.kiev.ua, rop@imath.kiev.ua\par

\vspace{2mm}\par\noindent {\it
$^\ddag$~Centre de Recherches Math\'ematiques,
Universit\'e de Montr\'eal,\\
$\phantom{^\ddag}$~C.P. 6128 succursale Centre-ville, Montr\'eal (Qu\'ebec), H3C 3J7 Canada}\\
$\phantom{^\ddag}$~E-mail: patera@CRM.UMontreal.CA
\par

\vspace{2mm}\par\noindent {\it
\noindent $^\S$~Wolfgang Pauli Institut, Oskar-Morgenstern-Platz 1, 1090 Wien, Austria\par
}

\vspace{5mm}\par\noindent\hspace*{10mm}\parbox{145mm}{\small
The invariants of solvable Lie algebras with nilradicals isomorphic to the algebra of strictly upper triangular matrices
and diagonal nilindependent elements are studied exhaustively.
Bases of the invariant sets of all such algebras are constructed by an original purely algebraic algorithm
based on Cartan's method of moving frames.
}\par\vspace{5mm}

\noindent
{\it Key words:} invariants of Lie algebras; Casimir operators; triangular matrices; moving frames.

\medskip

\noindent
{\it \it 2000 Mathematics Subject Classification:} 17B05; 17B10; 17B30; 22E70; 58D19; 81R05.
\noprint{
17-XX	Nonassociative rings and algebras
	17Bxx	Lie algebras and Lie superalgebras {For Lie groups, see 22Exx}
		17B05  	Structure theory
		17B08  	Coadjoint orbits; nilpotent varieties
		17B10  	Representations, algebraic theory (weights)
		17B30  	Solvable, nilpotent (super)algebras
22E70  	Applications of Lie groups to physics; explicit representations [See also 81R05, 81R10]
58D19  	Group actions and symmetry properties
81R05  	Finite-dimensional groups and algebras motivated by physics and their representations [See also 20C35, 22E70]
}

\section{Introduction}

The purpose of this paper is to present the advantages of our purely
algebraic algorithm for the construction of invariants with examples
of solvable Lie algebras with nilradicals isomorphic to the algebra
of strictly upper triangular matrices and nilindependent elements
represented by diagonal matrices. In contrast to known methods, this
approach is powerful enough to construct invariants of such Lie
algebras in a closed form. First let us present the motivation
behind this investigation.

Established work about invariants of Lie algebras can be
conditionally divided into two mainstream types that are weakly
connected with each other. One of them is more `physical' and is
mainly oriented to applications of invariants. The other one is more
`theoretical' and usually has a stronger mathematical background. We
simultaneously survey works on the invariants within the frameworks
of both. Note that invariant polynomials in Lie algebra elements are
called the \textit{Casimir operators}, while invariants that are not
necessarily polynomials are called \textit{generalized Casimir
operators}.

The term `Casimir operator' arose in the physical literature as a
reference to~\cite{Casimir}. At that time, only the lowest rank Lie
algebras appeared to be of interest. In subsequent years the need to
know the invariants of much larger Lie algebras arose more rapidly
in physics than in mathematics.

In the mathematics literature it was soon recognized that the
universal enveloping algebra ${\rm U}(\mathfrak{g})$ of a semisimple Lie
algebra $\mathfrak{g}$ contains elements (necessarily polynomial)
that commute with any element of $\mathfrak{g}$, that there is a
basis for all such invariants, and that the number of basis elements
coincides with the rank of $\mathfrak{g}$. The degrees of the basis
elements are given by the values of the exponents of the
corresponding Weyl group (augmented by~$1$).
The best known are the Casimir operators of degree $2$ for
semisimple Lie algebras. The explicit form of Casimir operators
depends on the choice of the basis of $\mathfrak{g}$. The center of
the universal enveloping algebra ${\rm U}(\mathfrak{g})$ proved to be
isomorphic to the space of polynomials on the dual space
to~$\mathfrak{g}$, which are invariant with respect to the
coadjoint action of the corresponding Lie group~\cite{Gelfand1950}.
This gives a basis for the calculation of Casimir operators by the
infinitesimal and algebraic methods.

There are numerous papers on the properties and the specific
computation of invariants of Lie algebras, on the estimation of
their number and on the application of invariants of various classes
of Lie algebras, or even of a particular Lie algebra which appears
in physical problems (see the citations of this paper and references
therein). Casimir operators are of fundamental importance in
physics. They represent such important quantities as angular
momentum, elementary particle mass and spin, Hamiltonians of various
physical systems and they also provide information on quantum
numbers that allow the characterization of the states of a system,
etc. Generalized Casimir operators of Lie algebras are of great
significance to representation theory as their eigenvalues provide
labels to distinguish irreducible representations. For this reason
it is of importance to have an effective procedure to determine
these invariants explicitly, in order to evaluate them for the
different representations of Lie algebras.

Unfortunately, up to the semi-simple case, which was completely
solved in the 1960's, there is no general theory that allows the
construction of the generalized Casimir operators of Lie algebras.
The standard infinitesimal method became conventional for the
calculations of invariants. It is based on integration of
overdetermined systems of first-order linear partial differential
equations associated with infinitesimal operators of coadjoint
action. This is why it is effective only for the algebras of a quite
simple structure or of low dimensions.

The interest in finding all independent invariants of Lie algebras
was recognized a few decades ago
\cite{Abellanas&MartinezAlonso1975,Beltrametti&Blasi1966,
Patera&Sharp&Winternitz1976,Pauri&Prosperi1966,Perelomov&Popov1968,Racah1950,Zassenhaus1977}.
In particular, functional bases of invariants were calculated for
all three-, \mbox{four-}, five-dimen\-sio\-nal and nilpotent
six-dimensional real Lie algebras
in~\cite{Patera&Sharp&Winternitz1976}. The same problem was
considered in~\cite{Ndogmo2000} for the six-dimensional real Lie
algebras with four-dimensional nilradicals.
In~\cite{Patera&Sharp&Winternitz1976a} the subgroups of the
Poincar\'e group along with their invariants were found. There is a
more detailed review of the low-dimensional algebras and their
invariants in
\cite{Boyko&Patera&Popovych2006,popovych&boyko&nesterenko&lutfullin2003}.
The cardinality of invariant bases was calculated by different
formulas within the framework of the infinitesimal
approach~\cite{Beltrametti&Blasi1966,Campoamor-Stursberg2004}.
Invariants of Lie algebras with various additional structural
restrictions were also constructed. Namely, the solvable Lie
algebras with the nilradicals isomorphic to the Heisenberg
algebras~\cite{Rubin&Winternitz1993}, with Abelian
nilradicals~\cite{Ndogmo2000a,Ndogmo&Wintenitz1994b}, with
nilradicals containing Abelian ideals of
codimension~1~\cite{Snobl&Winternitz2005}, solvable triangular
algebras~\cite{Tremblay&Winternitz2001}, some solvable rigid Lie
algebras~\cite{Campoamor-Stursberg2002a,Campoamor-Stursberg2003b},
solvable Lie algebras with graded nilradical of maximal nilindex and
a Heisenberg
subalgebra~\cite{Ancochea&Campoamor-Stursberg&GarciaVergnolle2006},
different classes of unsolvable
algebras~\cite{Campoamor-Stursberg2006a,Campoamor-Stursberg2007,Ndogmo2004}.
Empiric techniques were also applied for finding invariants of Lie
algebras (e.g. \cite{Barannyk&Fushchych1986}).

The existence of bases consisting entirely of Casimir operators
(polynomial invariants) is important for the theory of generalized
Casimir operators and for their applications. It was shown that it
is the case for the semi-simple, nilpotent, perfect and more general
algebraic Lie
algebras~\cite{Abellanas&MartinezAlonso1975,Abellanas&MartinezAlonso1979}.
Properties of Casimir operators of some perfect Lie algebras and
estimations for their number were investigated recently
in~\mbox{\cite{Campoamor-Stursberg2003c,Campoamor-Stursberg2003e,Ndogmo2004}}.

\looseness=-1
In~\cite{Boyko&Patera&Popovych2006,Boyko&Patera&Popovych2007,Boyko&Patera&Popovych2007b}
an original pure algebraic approach to invariants of Lie algebras
was proposed and developed. Within its framework, the technique of
Cartan's method of moving frames~\cite{Cartan1935,Cartan1937} in the
Fels--Olver version~\cite{Fels&Olver1998,Fels&Olver1999} is
specialized for the case of coadjoint action of the associated inner
automorphism groups on the dual spaces of Lie algebras. (For modern
development of the moving frames method and more references see also
\cite{Olver&Pohjanpelto2008}). Unlike the infinitesimal methods
based on solving systems of partial differential equations, such an
approach involves only systems of algebraic equations. As a result,
it is essentially simpler to extend the field of its application.
Note that similar algebraic tools were occasionally applied to
construct invariants for the specific case of inhomogeneous algebras
\cite{Kaneta1984a,Kaneta1984b,Perroud1983}. By the infinitesimal
method, such algebras were investigated
in~\cite{Chaichian&Demichev&Nelipa1983}.

Different versions of the algebraic approach were tested for the Lie
algebras of dimensions not greater than 6
\cite{Boyko&Patera&Popovych2006} and also a wide range of known
solvable Lie algebras of arbitrary finite dimensions with fixed
structure of nilradicals~\cite{Boyko&Patera&Popovych2007}. A special
technique for working with solvable Lie algebras having triangular
nilradicals was developed in~\cite{Boyko&Patera&Popovych2007b}.
Fundamental invariants were constructed with this technique for the
algebras $\mathfrak t_0(n)$, $\mathfrak t(n)$ and $\mathfrak{st}(n)$. 
Here $\mathfrak t_0(n)$ denotes the nilpotent
Lie algebra of strictly upper triangular $n\times n$ matrices over the field $\mathbb F$, 
where $\mathbb F$ is either $\mathbb C$ or $\mathbb R$. 
The solvable Lie algebras of non-strictly upper triangle and special upper triangle $n\times n$
matrices are denoted by $\mathfrak t(n)$ and $\mathfrak{st}(n)$, respectively.

The invariants of Lie algebras having triangular nilradicals were
first studied in~\cite{Tremblay&Winternitz2001}, by the
infinitesimal method. The claim about the Casimir operators of
$\mathfrak t_0(n)$ and the conjecture on the invariants of
$\mathfrak{st}(n)$ from~\cite{Tremblay&Winternitz2001} were
completely corroborated in~\cite{Boyko&Patera&Popovych2007b}.
Another conjecture was formulated  in~\cite{Tremblay&Winternitz2001}
on the invariants of solvable Lie algebras having $\mathfrak t_0(n)$
as their nilradicals and possessing a minimal (one) number of
nilindependent `diagonal' elements. It was completed and rigourously
proved in~\cite{Boyko&Patera&Popovych2007c}. Within the framework of
the infinitesimal approach, necessary calculations are too
cumbersome in these algebras even for small values of~$n$ that it
demanded the thorough mastery of the method, and probably led to
partial computational experiments and to the impossibility of
proving the conjectures for arbitrary values of~$n$.

In this paper, bases of the invariant sets of all the solvable Lie
algebras with nilradicals isomorphic to $\mathfrak t_0(n)$ and $s$
`diagonal' nilindependent elements are constructed for arbitrary
relevant values of~$n$ and $s$ (i.e., $n>1$, $0\leqslant s\leqslant
n-1$). We use the algebraic approach first proposed
in~\cite{Boyko&Patera&Popovych2006} along with some additional
technical tools developed for triangular and close algebras
in~\cite{Boyko&Patera&Popovych2007b,Boyko&Patera&Popovych2007c}. The
description of the necessary notions and statements, the precise
formulation and discussion of technical details of the applied
algorithm can be found ibid and are additionally reviewed in
Section~\ref{SectionAlgorithm} for convenience. 
In Section~\ref{SectionIllustrativeExample} an illustrative example on invariants of
a four-dimensional Lie algebra from the above class 
is given for clear demonstration of features of the developed method.

\looseness=-1
All the steps of the algorithm are implemented one after another for
the Lie algebras under consideration: construction of the coadjoint
representation of the corresponding Lie group and its fundamental
lifted invariant
(Section~\ref{SectionRepresentationOfCoadjointAction}), excluding
the group parameters from the lifted invariants by the normalization
procedure that results to a basis of the invariants for the
coadjoint action (Section~\ref{SectionInvariantsOfCoadjointAction})
and re-writing this basis as a basis of the invariants of the Lie
algebra under consideration (Section~\ref{SectionInvariants}). The
calculations for all steps are more complicated than
in~\cite{Boyko&Patera&Popovych2007b,Boyko&Patera&Popovych2007c}, but
due to optimization they remain quite useful. The necessary numbers
of normalization constraints, their forms and, therefore, the
cardinalities of the fundamental invariants depend on the algebra
parameters. In Section~\ref{SectionPartialCases} various particular
cases of the solvable Lie algebras with triangular nilradicals and
`diagonal' nilindependent elements, which was investigated earlier,
are connected with the obtained results.

\section{The algorithm}\label{SectionAlgorithm}

For convenience of the reader and to introduce some necessary
notations, before the description of the algorithm, we briefly
repeat the preliminaries given
in~\cite{Boyko&Patera&Popovych2006,Boyko&Patera&Popovych2007,Boyko&Patera&Popovych2007b} about
the statement of the problem of calculating Lie algebra invariants,
and on the implementation of the moving frame
method~\cite{Fels&Olver1998,Fels&Olver1999}. The comparative
analysis of the standard infinitesimal and the presented algebraic
methods, as well as their modifications, is given in~\cite{Boyko&Patera&Popovych2007b}.

Consider a Lie algebra~$\mathfrak g$ of dimension $\dim \mathfrak
g=n<\infty$ over the (complex or real) field~$\mathbb F$ and the corresponding
connected Lie group~$G$. Let~$\mathfrak g^*$ be the dual space of
the vector space~$\mathfrak g$. The map ${\rm Ad}^*\colon G\to
{\rm GL}(\mathfrak g^*)$, defined for each $g\in G$ by the relation
\[
\langle{\rm Ad}^*_g x,u\rangle=\langle x,{\rm Ad}_{g^{-1}}u\rangle
\quad \mbox{for all $x\in \mathfrak g^*$ and $u\in \mathfrak g$}
\]
is called the {\it coadjoint representation} of the Lie group~$G$.
Here ${\rm Ad}\colon G\to {\rm GL}(\mathfrak g)$ is the usual adjoint
representation of~$G$ in~$\mathfrak g$, and the image~${\rm Ad}_G$
of~$G$ under~${\rm Ad}$ is the inner automorphism group
of the Lie algebra~$\mathfrak g$.
The image of~$G$ under~${\rm Ad}^*$ is a subgroup of~${\rm GL}(\mathfrak g^*)$ and
is denoted by~${\rm Ad}^*_G$.

A smooth function $F\colon\Omega\to\mathbb F$, where $\Omega$ is a domain in~$\mathfrak g^*$, 
is called a (global in $\Omega$) {\it invariant} of~${\rm Ad}^*_G$  if
$
F({\rm Ad}_g^* x)=F(x)\ \mbox{for all}\ g\in G \ \mbox{and}\ x\in\Omega \ \mbox{such that}\ {\rm Ad}^*_g x\in\Omega.
$
The set of invariants of ${\rm Ad}^*_G$ on $\Omega$ is denoted by $\Inv({\rm Ad}^*_G)$
without an explicit indication of the domain~$\Omega$. 
Let below $\Omega$ is a neighborhood of a point from a regular orbit. 
It can always be chosen in such a way that the group~${\rm Ad}^*_G$ acts regularly on $\Omega$. 
Then the maximal number $N_\mathfrak g$ of functionally independent invariants in 
$\Inv({\rm Ad}^*_G)$ coincides with the codimension of the regular orbits of~${\rm Ad}^*_G$, 
i.e., it is given by the difference
\[
N_\mathfrak g=\dim \mathfrak g-\rank {\rm Ad}^*_G.
\]
Here $\rank {\rm Ad}^*_G$ denotes the dimension of the regular
orbits of~${\rm Ad}^*_G$ and will be called the {\em rank of the
coadjoint representation} of~$G$ (and of~$\mathfrak g$). It is a
basis independent characteristic of the algebra~$\mathfrak g$, the
same as $\dim \mathfrak g$ and $N_\mathfrak g$.

To calculate the invariants explicitly, one should fix a basis
$\mathcal E=(e_1,\ldots,e_n)$ of the algebra~$\mathfrak g$. It
leads to fixing the dual basis $\mathcal E^*=(e_1^*,\ldots,e_n^*)$
in the dual space~$\mathfrak g^*$ and to the identification of
${\rm Ad}_G$ and ${\rm Ad}^*_G$ with the associated matrix
groups. The basis elements $e_1,\ldots,e_n$ satisfy the commutation
relations $[e_i,e_j]=\sum_{k=1}^nc_{ij}^k e_k$, $i,j=1,\ldots,n$,
where $c_{ij}^k$ are components of the tensor of structure constants
of~$\mathfrak g$ in the basis~$\mathcal E$.

Let $x\to\check x=(x_1,\ldots,x_n)$ be the (local) coordinates in~$\mathfrak
g^*$ associated with~$\mathcal E^*$. Given any invariant
$F(x_1,\ldots,x_n)$ of~${\rm Ad}^*_G$, one finds the corresponding
invariant of the Lie algebra~$\mathfrak g$ by symmetrization,
$\mathop{\rm Sym}\nolimits F(e_1,\ldots,e_n)$, of $F$. It is often
called a \emph{generalized Casimir operator} of~$\mathfrak g$. If
$F$ is a~polynomial, $\mathop{\rm Sym}\nolimits F(e_1,\ldots,e_n)$
is a usual \emph{Casimir operator}, i.e., an element of the center of
the universal enveloping algebra of~$\mathfrak g$. More precisely,
the symmetrization operator~$\mathop{\rm Sym}\nolimits$ acts only on
the monomials of the forms~$e_{i_1}\cdots e_{i_r}$, where there are
non-commuting elements among~$e_{i_1}, \ldots, e_{i_r}$, and is
defined by the formula
\[
\mathop{\rm Sym}\nolimits (e_{i_1}\cdots e_{i_r})=\dfrac1{r!}\sum_{\sigma\in{\rm S}_r}e_{i_{\sigma_1}}\cdots e_{i_{\sigma_r}},
\]
where $i_1, \ldots, i_r$ take values from 1 to $n$, $r\geqslant 2$.
The symbol ${\rm S}_r$ denotes the symmetric group on $r$~letters. 
The set of invariants of $\mathfrak g$ is denoted by
$\Inv(\mathfrak g)$.

A set of functionally independent invariants $F^l(x_1,\ldots,x_n)$,
\mbox{$l=1,\ldots,N_\mathfrak g$}, forms {\it a~functional basis}
({\it fundamental invariant}) of $\Inv({\rm Ad}^*_G)$, i.e., 
each invariant $F(x_1,\ldots,x_n)$ can be uniquely represented as
a~function of~$F^l(x_1,\ldots,x_n)$, \mbox{$l=1,\ldots,N_\mathfrak
g$}. Accordingly the set of $\mathop{\rm Sym}\nolimits
F^l(e_1,\ldots,e_n)$, \mbox{$l=1,\ldots,N_\mathfrak g$}, is called a
basis of~$\Inv(\mathfrak g)$.

Our task here is to determine the basis of the functionally
independent invariants for ${\rm Ad}^*_G$, and then to transform
these invariants into the invariants of the algebra~$\mathfrak g$.
Any other invariant of $\mathfrak g$ is a function of the
independent ones.

Let us recall some facts from \cite{Fels&Olver1998,Fels&Olver1999}
and adapt them to the particular case of the coadjoint action of~$G$
on~$\mathfrak g^*$. Let~$\mathcal G={\rm Ad}^*_G\times \mathfrak
g^*$ denote the trivial left principal ${\rm Ad}^*_G$-bundle
over~$\mathfrak g^*$. The right regularization~$\widehat R$ of the
coadjoint action of~$G$ on~$\mathfrak g^*$ is the diagonal action
of~${\rm Ad}^*_G$ on~$\mathcal G={\rm Ad}^*_G\times \mathfrak g^*$.
It is provided by the map
$
\widehat R_g({\rm Ad}^*_h,x)=({\rm Ad}^*_h\cdot {\rm Ad}^*_{g^{-1}},{\rm Ad}^*_g x),
\ g,h\in G, \ x\in \mathfrak g^*,
$
where the action on the bundle~$\mathcal G={\rm Ad}^*_G\times
\mathfrak g^*$ is regular and free. We call \raisebox{0ex}[0ex][0ex]{$\widehat R_g$} the
\emph{lifted coadjoint action} of~$G$. It projects back to the
coadjoint action on~$\mathfrak g^*$ via the ${\rm
Ad}^*_G$-equivariant projection~$\pi_{\mathfrak g^*}\colon
\mathcal G\to \mathfrak g^*$. Any \emph{lifted invariant} of~${\rm
Ad}^*_G$ is a (locally defined) smooth function from~$\mathcal G$
to a~manifold, which is invariant with respect to the lifted
coadjoint action of~$G$. The function $\mathcal
I\colon\mathcal G\to \mathfrak g^*$ given by $\mathcal I=\mathcal
I({\rm Ad}^*_g,x)={\rm Ad}^*_g x$ is the \emph{fundamental lifted
invariant} of ${\rm Ad}^*_G$, i.e., $\mathcal I$ is a lifted
invariant, and each lifted invariant can be locally written as a
function of~$\mathcal I$. Using an arbitrary function~$F(x)$
on~$\mathfrak g^*$, we can produce the lifted
invariant~$F\circ\mathcal I$ of~${\rm Ad}^*_G$ by replacing $x$ with
$\mathcal I={\rm Ad}^*_g x$ in the expression for~$F$. Ordinary
invariants are particular cases of lifted invariants, where one
identifies any invariant formed as its composition with the standard
projection~$\pi_{\mathfrak g^*}$. Therefore, ordinary invariants are
particular functional combinations of lifted ones that happen to be
independent of the group parameters of~${\rm Ad}^*_G$.

The \emph{algebraic algorithm} for finding invariants of the Lie
algebra $\mathfrak g$ is briefly formulated in the following four
steps.

\medskip

1. {\it Construction of the generic matrix $B(\theta)$ of~${\rm Ad}^*_G$.} 
$B(\theta)$ is the matrix of an inner automorphism of the
Lie algebra~$\mathfrak g$ in the given basis $e_1$, \ldots, $e_n$,
$\theta=(\theta_1,\ldots,\theta_r)$ is a complete tuple of group
parameters (coordinates) of~${\rm Ad}_G$, and
$r=\dim{\rm Ad}^*_G=\dim{\rm Ad}_G=n-\dim{\rm Z}(\mathfrak g),$ 
where ${\rm Z}(\mathfrak g)$ is the center of~$\mathfrak g$.

\medskip

2. {\it Representation of the fundamental lifted invariant.} The
explicit form of the fundamental lifted invariant~$\mathcal
I=(\mathcal I_1,\ldots,\mathcal I_n)$ of ${\rm Ad}^*_G$ in the
chosen coordinates~$(\theta,\check x)$ in ${\rm Ad}^*_G\times\mathfrak g^*$ is $\mathcal I=\check x\cdot B(\theta)$, i.e.,
$(\mathcal I_1,\ldots,\mathcal I_n)=(x_1,\ldots,x_n)\cdot B(\theta_1,\ldots,\theta_r)$.

\medskip

3. {\it Elimination of parameters by normalization}. 
We choose the maximum possible number $\rho$ of lifted invariants $\mathcal
I_{j_1}$, \ldots, $\mathcal I_{j_\rho}$, constants $c_1$, \ldots, $c_\rho$ and group parameters
$\theta_{k_1}$,~\ldots,~$\theta_{k_\rho}$ such that the equations
$\mathcal I_{j_1}=c_1$, \ldots, $\mathcal I_{j_\rho}=c_\rho$ are
solvable with respect to $\theta_{k_1}$,~\ldots,~$\theta_{k_\rho}$.
After substituting the found values of
$\theta_{k_1}$,~\ldots,~$\theta_{k_\rho}$ into the other lifted
invariants, we obtain $N_\mathfrak g=n-\rho$ expressions $F^l
(x_1,\ldots,x_n)$ without $\theta$'s.

\medskip

4. {\it Symmetrization.} 
The functions $F^l(x_1,\ldots,x_n)$ necessarily form a basis of~$\Inv({\rm Ad}^*_G)$. 
They are symmetrized to $\mathop{\rm Sym}\nolimits F^l(e_1,\ldots,e_n)$. 
It is the desired basis of~$\Inv(\mathfrak g)$.

\medskip

Following the preceding papers
\cite{Boyko&Patera&Popovych2007b,Boyko&Patera&Popovych2007c}
on invariants of the triangular Lie algebras,
here we use, in contrast with the general situation,
special coordinates for inner automorphism groups,
which naturally harmonize with the canonical matrix
representations of the corresponding Lie groups and with special
`matrix' enumeration of a part of the basis elements.
The individual approach results in the clarification and a substantial
reduction of all calculations. Thus, algebraic systems
solved under normalization are reduced to linear ones.

The essence of the normalization procedure by Fels and Olver
\cite{Fels&Olver1998,Fels&Olver1999} can be presented in the form of
on the following statement~\cite{Boyko&Patera&Popovych2007b}.

\begin{proposition}\label{PropositionOnNormalization}
Let~$\mathcal I=(\mathcal I_1,\ldots,\mathcal I_n)$ be a fundamental
lifted invariant of ${\rm Ad}^*_G$, for the lifted invariants $\mathcal I_{j_1}$,
\ldots, $\mathcal I_{j_\rho}$ and some constants $c_1$, \ldots,
$c_\rho$ the system $\mathcal I_{j_1}=c_1$, \ldots, $\mathcal
I_{j_\rho}=c_\rho$ be solvable with respect to the parameters
$\theta_{k_1}$,~\ldots,~$\theta_{k_\rho}$ and substitution of the
found values of $\theta_{k_1}$,~\ldots,~$\theta_{k_\rho}$ into the
other lifted invariants result in $m=n-\rho$ expressions
$\hat{\mathcal I}_l$, $l=1,\dots,m$, depending only on $x$'s. Then
$\rho=\rank {\rm Ad}^*_G$, $m=N_\mathfrak g$ and $\hat{\mathcal
I}_1$, \ldots, $\hat{\mathcal I}_m$ form a basis of $\Inv({\rm
Ad}^*_G)$.
\end{proposition}

Our experience on the calculation of invariants of a wide range of
Lie algebras shows that the version of the algebraic method, which
is based on Proposition~\ref{PropositionOnNormalization}, is most
effective.
In particular, it provides finding the cardinality of the invariant
basis in the process of construction of the invariants.
It is the version that is used in this paper.

\section{Illustrative example}\label{SectionIllustrativeExample}

Before the calculation of invariants for the general case of Lie algebras from the class under consideration, 
we present an illustrative example on invariants of a low-dimensional Lie algebra from this class. 
This demonstrates features of the developed method. 

The four-dimensional solvable Lie algebra~$\mathfrak{g}_{4.8}^{b}$ has the following nonzero commutation relations
\begin{gather*}
[e_2,e_3]=e_1,  \quad [e_1,e_4]=(1+b)e_1, \quad 
[e_2,e_4]=e_2 , \quad [e_3,e_4]=be_3, \quad |b|\leq 1.
\end{gather*}
Its nilradical is three-dimensional and isomorphic to the Weil--Heisenberg algebra $\mathfrak{g}_{3.1}$. 
(Here we use the notations of low-dimensional Lie algebras according to Mubarakzyanov's classification~\cite{mubarakzyanov1963.1}.)

We construct a presentation of the inner automorphism matrix $B(\theta)$ of the Lie algebra~$\mathfrak g$, involving 
second canonical coordinates on ${\rm Ad}_G$ as group parameters~$\theta$ 
\cite{Boyko&Patera&Popovych2006,Boyko&Patera&Popovych2007,Boyko&Patera&Popovych2007b}.
The matrices~$\hat{\rm ad}_{e_i}$, $i=1,\dots,4,$ of the adjoint representation 
of the basis elements $e_1$, $e_2$, $e_3$ and $e_4$ respectively have the form
\begin{gather*}
\left(\begin{array}{cccc} 0& 0 & 0 & 1+b \\ 0&0&0&0\\0&0&0&0\\ 0&0&0&0 \end{array}\right)\!, \ 
\left(\begin{array}{cccc} 0& 0 & 1 & 0 \\ 0&0&0&1\\0&0&0&0\\ 0&0&0&0 \end{array}\right)\!, \ 
\left(\begin{array}{cccc} 0& -1 & 0 & 0 \\ 0&0&0&0\\0&0&0&b\\ 0&0&0&0 \end{array}\right)\!, \ 
\left(\begin{array}{cccc} -1-b& 0 & 0 & 0 \\ 0&-1&0&0\\0&0&-b&0\\ 0&0&0&0 \end{array}\right)\!.
\end{gather*}
The inner automorphisms of~$\mathfrak{g}_{4.8}^{b}$ are then described by the triangular matrix
\begin{gather*}
B(\theta)=\prod_{i=1}^3\exp(\theta_i\hat{\rm ad}_{e_i}) \cdot \exp(-\theta_4\hat{\rm ad}_{e_4})
=\left(\begin {array}{@{}c@{\,\,\,}c@{\,\,\,}c@{\,\,\,}c@{}}
e^{(1+b)\theta_4} & -\theta_3 e^{\theta_4} & \theta_2 e^{b \theta_4}&
b\theta_2\theta_3+(1+b)\theta_1\\
0&e^{\theta_4}  & 0 &\theta_2 \\
0& 0 &e^{b \theta_4}   & b\theta_3 \\
0&0&0&1
\end {array}\right)\!.
\end{gather*}
Therefore, a functional basis of lifted invariants is formed by
\begin{gather*}
\mathcal{I}_1=e^{(1+b)\theta_4} x_1,\\
\mathcal{I}_2=e^{\theta_4}(-\theta_3  x_1+ x_2),\\
\mathcal{I}_3=e^{b\theta_4} (\theta_2 x_1 + x_3),\\
\mathcal{I}_4=(b\theta_2\theta_3+(1+b)\theta_1)x_1+\theta_2x_2+b\theta_3x_3+x_4.
\end{gather*}

Further the cases $b=-1$ and $b\not=-1$ should be considered separately.

There are no invariants in the case $b\not=-1$ since in view of 
Proposition~\ref{PropositionOnNormalization} the number of functionally independent invariants is equal to zero.
Indeed, the system $\mathcal I_1=1$, $\mathcal I_2=\mathcal I_3=\mathcal I_4=0$ 
is solvable with respect to the whole set of the parameters~$\theta$. 

It is obvious  that in the case $b=-1$ the element $e_1$ generating the center~${\rm Z}(\mathfrak{g}_{4.8}^{-1})$ is an invariant.
(The corresponding lifted invariant $\mathcal{I}_1=x_1$ does not depend on the parameters~$\theta$.)
Another invariant is easily found via combining the lifted invariants:
$\mathcal{I}_1\mathcal{I}_4-\mathcal{I}_2\mathcal{I}_3=x_1x_4-x_2x_3$.
After the symmetrization procedure we obtain the following polynomial basis of the invariant set of this algebra
\[
e_1, \quad e_1e_4-\frac{e_2e_3+e_3e_2}{2}.
\]
The second basis invariant can be also constructed by the normalization technique. 
We solve the equations $\mathcal I_2=\mathcal I_3=0$ with respect to the parameters $\theta_2$ and $\theta_3$ and 
substitute the expressions for them into the lifted invariant $\mathcal{I}_4$. 
The obtained expression $x_4-x_2x_3/x_1$ does not contain the parameters~$\theta$ and, therefore, is an invariant 
of the coadjoint representation. 
For the basis of invariants to be polynomial, we multiply this invariant by the invariant~$x_1$. 
It is the technique that is applied below for the general case of the Lie algebras under consideration.

Note that in the above example the symmetrization procedure can be assumed trivial 
since the symmetrized invariant $e_1e_4-\frac12(e_2e_3+e_3e_2)$ differs from the non-symmetrized version 
$e_1e_4-e_2e_3$ (resp. $e_1e_4-e_3e_2$) on the invariant $\frac12e_1$ (resp. $-\frac12e_1$). 
If we take the rational invariant $e_4-e_2e_3/e_1$ (resp. $e_4-e_3e_2/e_1$), 
the symmetrization is equivalent to the addition of the constant~$\frac12$ (resp. $-\frac12$). 

Invariants of~$\mathfrak{g}_{4.8}^b$ were first described in~\cite{Patera&Sharp&Winternitz1976} within the framework 
of the infinitesimal approach.

\section{Structure of algebras}\label{SectionAlgebraStructure}

Consider the solvable Lie algebra $\mathfrak t_\gamma(n)$
with the nilradical $\NilRad(\mathfrak t_\gamma(n))$ isomorphic to $\mathfrak t_0(n)$ and
$s$ nilindependent element $f_p$, $p=1,\ldots,s$, which act on elements of the nilradical in the way
as the diagonal matrices $\Gamma_p=\diag(\gamma_{p1},\dots,\gamma_{pn})$ act on strictly upper triangular matrices.
The matrices $\Gamma_p$, $p=1,\ldots,s$, and the identity matrix are jointly linear independent 
since otherwise \mbox{$\NilRad(\mathfrak t_\gamma(n))\ne\mathfrak t_0(n)$}.
Therefore, the algebra $\mathfrak t_\gamma(n)$ is naturally embedded into $\mathfrak t(n)$ as an ideal
under identification of $\NilRad(\mathfrak t_\gamma(n))$ with $\mathfrak t_0(n)$ and of $f_p$ with $\Gamma_p$.

We choose the concatenation of the canonical basis of $\NilRad(\mathfrak t_\gamma(n))$ and of the $s$-element tuple $(f_p,\,p=1,\ldots,s)$
as the canonical basis of $\mathfrak t_\gamma(n)$.
In the basis of $\NilRad(\mathfrak t_\gamma(n))$ we use `matrix' enumeration of basis elements
$e_{ij}$, $i<j$, with the `increasing' pair of indices similarly to
the canonical basis $(E^n_{ij},\,i<j)$ of the isomorphic matrix algebra $\mathfrak t_0(n)$.

Hereafter
$E^n_{ij}$ (for the fixed values $i$ and $j$) denotes the $n\times n$ matrix $(\delta_{ii'}\delta_{jj'})$
with $i'$ and $j'$ running the numbers of rows and columns, respectively,
i.e., the $n\times n$ matrix with the unit on the cross of the $i$th row and the $j$th column and the zero otherwise.
The indices $i$, $j$, $k$ and $l$ run at most from~1 to~$n$.
Only additional constraints on the indices are indicated.
The subscripts $p$ and~$p'$ run from~1 to~$s$, the subscripts $q$ and~$q'$ run from~1 to~$s'$.
The summation convention over repeated indices $p$, $p'$, $q$ and $q'$ is used unless otherwise stated.
The number~$s$ is in the range $0,\dots,n-1$.
In the case $s=0$ we assume $\gamma=0$, and all terms with the subscript~$p$ should be omitted from consideration.
The value $s'$ ($s'<s$) is defined in Proposition~\ref{PropositionOnReducedFormOfParameterMatrix} below.

Thus, the basis elements $e_{ij}\sim E^n_{ij}$, $i<j$, and $f_p\sim\sum_i \gamma_{pi}E^n_{ii}$ satisfy the commutation relations
\[
[e_{ij},e_{i'\!j'}]=\delta_{i'\!j}e_{ij'}-\delta_{ij'}e_{i'\!j}, \quad
[f_p,e_{ij}]=(\gamma_{pi}-\gamma_{pj})e_{ij}, \quad
\]
where $\delta_{ij}$ is the Kronecker delta.

The Lie algebra $\mathfrak t_\gamma(n)$ can be considered as the Lie algebra of the Lie subgroup
\[
{\rm T}_\gamma(n)=\{B\in {\rm T}(n)\mid \exists \, \varepsilon_p\in\mathbb F\colon b_{ii}=e^{\gamma_{pi}\varepsilon_p}\}
\]
of the Lie group ${\rm T}(n)$ of non-singular upper triangular $n\times n$ matrices.

\begin{proposition}\label{PropositionOnIsomorphismsOft_gamma(n)'s}
The algebras $\mathfrak t_\gamma(n)$ and $\mathfrak t_{\gamma'}(n)$ are isomorphic if  and only if 
there exist $\lambda\in{\rm GL}(s,\mathbb F)$ and  $\mu\in\mathbb F^s$ such that either 
\begin{gather*}
\gamma'_{pi}=\lambda_{pp'}\gamma_{p'\!i}+\mu_p,\quad p=1,\dots,s,\ i=1,\dots,n,
\\
\hspace*{-\mathindent}\mbox{or}\\
\gamma'_{pi}=\lambda_{pp'}\gamma_{p'\!,n-i+1}+\mu_p,\quad p=1,\dots,s,\ i=1,\dots,n.
\end{gather*}
\end{proposition}

\begin{proof}
Embedding the algebras $\mathfrak t_\gamma(n)$ and $\mathfrak t_{\gamma'}(n)$ into $\mathfrak t(n)$, 
we obtain that the nilradicals of these two algebras coincide, 
$\mathfrak n:=\NilRad(\mathfrak t_\gamma(n))=\NilRad(\mathfrak t_{\gamma'}(n))=\mathfrak t_0(n)$. 
By the construction,  have the same canonical basis of the common nilradical.
For each~$p$, denote by $f'_p$ the basis element of $\mathfrak t_{\gamma'}(n)$ 
that is associated with the diagonal matrices $\Gamma'_p=\diag(\gamma'_{p1},\dots,\gamma'_{pn})$. 

Suppose that the algebras $\mathfrak t_\gamma(n)$ and $\mathfrak t_{\gamma'}(n)$ are isomorphic. 
Let $\varphi\colon\mathfrak t_{\gamma'}(n)\to\mathfrak t_\gamma(n)$ be an isomorphism between these algebras. 
Since $\mathfrak n$ is the common nilradical of $\mathfrak t_\gamma(n)$ and $\mathfrak t_{\gamma'}(n)$, 
it is preserved by~$\varphi$, 
and thus the restriction of~$\varphi$ to $\mathfrak n$ is an automorphism of~$\mathfrak n=\mathfrak t_0(n)$. 
The automorphism group of the algebra~$\mathfrak t_0(n)$ is well-known; 
see, e.g.,~\cite{Cao&Tan2003} and references therein. 
These facts jointly give that $\varphi(f'_p)=\lambda_{pp'}f'_{p'}+O(\mathfrak n)$ for some nondegenerate matrix $(\lambda_{pp'})$
and either $\varphi(e_{ij})=e_{ij}+O(\mathfrak n^{j-i+1})$ for any $(i,j)$ with $i<j$ 
or $\varphi(e_{ij})=-e_{n-j+1,n-i+1}+O(\mathfrak n^{j-i+1})$ for any $(i,j)$ with $i<j$. 
Here $O(\mathfrak n^l)$ denotes an element of the $l$th megaideal~$\mathfrak n^l$ in the lower (descending) central series of~$\mathfrak n$,
$\mathfrak n^1:=\mathfrak n$, $\mathfrak n^l:=[\mathfrak n, \mathfrak n^{l-1}]$ for $l>1$.
For the first case for possible values of~$\varphi(e_{ij})$, we have 
\begin{gather*}
\varphi([f'_p,e_{ij}])=\varphi((\gamma'_{pi}-\gamma'_{pj})e_{ij})=(\gamma'_{pi}-\gamma'_{pj})e_{ij}+O(\mathfrak n^{j-i+1})
\\[-.9ex]\hspace{.3ex}\rule{.1ex}{1.5ex}\hspace{.3ex}\rule{.1ex}{1.5ex}\\[-.9ex]
[\varphi(f'_p),\varphi(e_{ij})])=[\lambda_{pp'}f_{p'}+O(\mathfrak n),e_{ij}+O(\mathfrak n^{j-i+1})]
=\lambda_{pp'}(\gamma_{p'\!i}-\gamma_{p'\!j})e_{ij}+O(\mathfrak n^{j-i+1}).
\end{gather*}
Therefore, $\gamma'_{pi}-\gamma'_{pj}=\lambda_{pp'}(\gamma_{p'\!i}-\gamma_{p'\!j})$, and thus
\[
\gamma'_{pi}-\lambda_{pp'}\gamma_{p'\!i}=\gamma'_{pj}-\lambda_{pp'}\gamma_{p'\!j}:=\mu_p,
\] 
which gives the first relation between~$\gamma$'s and~$\gamma'$'s from the proposition. 
The second case for possible values of~$\varphi(e_{ij})$ is considered in the similar way.

The sufficiency of the relations between~$\gamma$'s and~$\gamma'$'s 
for $\mathfrak t_\gamma(n)$ and $\mathfrak t_{\gamma'}(n)$ to be isomorphic is obvious.
\end{proof}

The parameter matrix $\gamma$ and $\gamma'$ are assumed equivalent 
if the algebras $\mathfrak t_\gamma(n)$ and $\mathfrak t_{\gamma'}(n)$ are isomorphic.
In~other words, the parameter matrix $\gamma=(\gamma_{pi})$ is defined up to 
nonsingular $s\times s$ matrix multiplier, 
entry summands that are homogeneous in rows, 
and the mirror reflection with respect to the central vertical line of~$\gamma$.
Up to the equivalence the additional condition $\mathop{\rm tr}\Gamma_p=\sum_{i}\gamma_{pi}=0$ can be imposed on the algebra parameters.
Therefore, the algebra $\mathfrak t_\gamma(n)$ can in fact be embedded, as an ideal, 
into $\mathfrak{st}(n)$ instead of~$\mathfrak t(n)$.

\begin{proposition}\label{PropositionOnReducedFormOfParameterMatrix}
Up to the equivalence relation on algebra parameters, the following conditions can be assumed satisfied
\begin{gather*}
\exists \, s'\in\left\{0,\dots,\min\left(s,\left[\frac n2\right]\right)\right\},\quad
\exists \, k_q,\ q=1,\dots,s',\quad 1\leqslant k_1<k_2<\dots<k_{s'\!}\leqslant\left[\frac n2\right]\colon
\\
\gamma_{qk}=\gamma_{q\varkappa},\ k<k_q,\quad
\gamma_{q\varkappa_q}-\gamma_{qk_q}=1,\quad
\gamma_{pk_q}=\gamma_{p\varkappa_q},\ p\ne q,\quad q=1,\dots,s', 
\\
\gamma_{pk}=\gamma_{p\varkappa},\ p>s',\ k=1, \dots, \left[\frac n2\right],
\end{gather*}
where $\varkappa:=n-k+1$, $\varkappa_q:=n-k_q+1$.
\end{proposition}

\begin{proof}
If $\gamma_{pk}=\gamma_{p\varkappa}$ for all $k\in\{1,\dots,[n/2]\}$ and all $p\in\{1,\dots,s\}$, then we set $s'=0$.
Otherwise, we set $k_1$ equal to the minimal value of~$k$ for which there exists $p_1$ such that $\gamma_{p_1k}\ne\gamma_{p_1\varkappa}$.
Permuting, scaling and combining rows of the matrix~$\gamma$, we make
$p_1=1$, $\gamma_{1\varkappa_1}-\gamma_{1k_1}=1$ and $\gamma_{pk_1}=\gamma_{p\varkappa_1}$, $p\ne 1$
that gives the conditions corresponding to $q=1$.

Then, if $\gamma_{pk}=\gamma_{p\varkappa}$ for all $k\in\{1,\dots,[n/2]\}$ and all $p\in\{2,\dots,s\}$, then we get $s'=1$.
Otherwise, we set $k_2$ equal to the minimal value of~$k$ for which there exists $p_2>p_1=1$ such that $\gamma_{p_2k}\ne\gamma_{p_2\varkappa}$.
It follows from the previous step that $k_2>k_1$.
Permuting, scaling and combining rows of the matrix~$\gamma$, we make
$p_2=2$, $\gamma_{2\varkappa_2}-\gamma_{2k_2}=1$ and $\gamma_{pk_2}=\gamma_{p\varkappa_2}$, $p\ne 2$.

By induction, iteration of this procedure leads to the statement.
\end{proof}

We will say that the parameter matrix~$\gamma$ is of a \emph{reduced form}
if it satisfies the conditions of Proposition~\ref{PropositionOnReducedFormOfParameterMatrix}. 
The equalities with $p>s'$ for~$\gamma$ in reduced form are in fact satisfied for all~$k$'s, 
$\gamma_{pk}=\gamma_{p\varkappa}$ with $p>s'$ and $k=1,\dots,n$.
Note also that
\[
s'=\rank(\gamma_{p\varkappa}-\gamma_{pk})^{p=1,\dots,s}_{k=1,\dots,[n/2]}
=\rank(\gamma_{p\varkappa}-\gamma_{pk})^{p=1,\dots,s}_{k=1,\dots,n}.
\]

\section{Representation of the coadjoint action}\label{SectionRepresentationOfCoadjointAction}

Let $e_{ji}^*$, $x_{ji}$ and~$y_{ij}$ denote
the basis element and the coordinate function in the dual space~$\mathfrak t_\gamma^*(n)$ and
the coordinate function in~$\mathfrak t_\gamma(n)$,
which correspond to the basis element~$e_{ij}$, $i<j$.
In~particular, $\langle e_{j'\!i'\!}^*,e_{ij}\rangle=\delta_{ii'\!}\delta_{jj'}.$
The reverse order of subscripts of the objects
associated with the dual space~$\mathfrak t_\gamma^*(n)$
is natural (see, e.g., \cite[Section 1.4]{Perelomov1990})
and additionally justified by the simplification of a matrix representation of lifted invariants.
$f_p^*$, $x_{p0}$ and $y_{p0}$ denote similar objects corresponding to the basis element~$f_p$.
We additionally set $y_{ii}=\gamma_{pi}y_{p0}$ 
and then complete the collections of~$x_{ji}$ and of~$y_{ij}$ to the matrices~ $X$ and~$Y$ with zeros.
Hence $X$ is a strictly lower triangular matrix and $Y$ is a non-strictly upper triangular one.
The analogous `matrix' whose $(i,j)$th entry is equal to $e_{ij}$ for $i<j$
and 0 otherwise is denoted by $\mathcal E$.

\begin{lemma}\label{LemmaOnLiftedInvsOfDiagSolvAlgsWithTriangularNilradical}
A complete set of functionally independent lifted invariants of ${\rm Ad}^*_{{\rm T}_\gamma(n)}$
is exhausted by the expressions
\[
\mathcal I_{ij}=\sum_{i\leqslant i',\,j'\leqslant j}b_{ii'}\widehat b_{j'\!j}x_{i'\!j'}, \quad j<i,
\qquad
\mathcal I_{p0}=x_{p0}+\sum_{j<i}\,\sum_{j\leqslant l\leqslant i}\gamma_{pl}b_{li}\widehat b_{jl}x_{ij},
\]
where
$B=(b_{ij})$ is an arbitrary matrix from ${\rm T}_\gamma(n)$,
$B^{-1}=(\widehat b_{ij})$ is the inverse matrix of~$B$.
\end{lemma}

\begin{proof} The adjoint action of $B\in{\rm T}_\gamma(n)$ on the matrix~$Y$ is
${\rm Ad}_BY=BYB^{-1}$, i.e.,
\[
{\rm Ad}_B\biggl(y_{p0}f_p+\sum_{i<j}y_{ij}e_{ij}\biggr)
=y_{p0}f_p+y_{p0}\sum_{i<j}\,\sum_{i\leqslant i'\leqslant j}b_{ii'}\gamma_{pi'\!}\widehat b_{i'\!j}e_{ij}
+\sum_{i\leqslant i'<j'\leqslant j}b_{ii'}y_{i'\!j'}\widehat b_{j'\!j}e_{ij}.
\]
After changing $e_{ij}\to x_{ji}$, $y_{ij}\to e_{ji}^*$, $f_p\to x_{p0}$, $y_{p0}\to f_p^*$,
$b_{ij}\leftrightarrow \widehat b_{ij}$
in the latter equality, we obtain the representation for the coadjoint action of~$B$
\begin{gather*}
{\rm Ad}_B^*\biggl(x_{p0}f_p^*+\sum_{i<j}x_{ji}e_{ji}^*\biggr)
=x_{p0}f_p^*+\sum_{i<j}\,\sum_{i\leqslant i'\leqslant j}b_{i'\!j}x_{ji}\widehat b_{ii'}\gamma_{pi'\!}f_p^*
+\sum_{i\leqslant i'<j'\leqslant j}b_{j'\!j}x_{ji}\widehat b_{ii'}e_{j'\!i'}^*
\\ \qquad
=\biggl(x_{p0}+\sum_{i<j}\,\sum_{i\leqslant i'\leqslant j}b_{i'\!j}x_{ji}\widehat b_{ii'}\gamma_{pi'\!}\biggr)f_p^*
+\sum_{i'<j'}(BXB^{-1})_{j'\!i'}e_{j'\!i'}^*.
\end{gather*}
Therefore, $\mathcal I_{p0}$ and the elements $\mathcal I_{ij}$, $j<i$, of the matrix $\mathcal I=BXB^{-1}$, where $B\in{\rm T}_\gamma(n)$,
form a fundamental lifted invariant of ${\rm Ad}^*_{{\rm T}_\gamma(n)}$.
\end{proof}

\begin{remark}
The complete set of parameters in the above representation of lifted invariants is formed by $b_{ij}$, $j<i$, and $\varepsilon_p$.
The center of the group ${\rm T}_\gamma(n)$ is nontrivial only if $\gamma_{p1}=\gamma_{pn}$, namely, then 
${\rm Z}({\rm T}_\gamma(n))=\{E^n+b_{1n}E^n_{1n},\ b_{1n}\in\mathbb F\}.$ 
Here $E^n=\diag(1,\ldots,1)$ is the $n\times n$ identity matrix.
In this case, the inner automorphism group of~$\mathfrak t_\gamma(n)$ is isomorphic to the factor-group
${\rm T}_\gamma(n)/{\rm Z}({\rm T}_\gamma(n))$ and hence its dimension is $\frac12n(n-1)$.
Then the parameter $b_{1n}$ in the representation of lifted invariants is inessential.
Otherwise, the inner automorphism group of~$\mathfrak t_\gamma(n)$ is isomorphic to the whole group ${\rm T}_\gamma(n)$
and all the parameters in the constructed lifted invariants are essential.
\end{remark}

\section{Invariants of the coadjoint action}\label{SectionInvariantsOfCoadjointAction}

Below $A^{i_1,i_2}_{j_1,j_2}$, where $i_1\leqslant i_2$, $j_1\leqslant j_2$,
denotes the submatrix $(a_{ij})^{i=i_1,\ldots,i_2}_{j=j_1,\ldots,j_2}$ of a matrix $A=(a_{ij})$.
The standard notation $|A|=\det A$ is used.
The conjugate values of $k$'s with respect to $n$ is denoted by respective $\varkappa$'s, i.e., 
\[
\varkappa:=n-k+1, \quad \varkappa_q:=n-k_q+1,\quad \varkappa':=n-k'+1.
\]

Similarly to~\cite{Boyko&Patera&Popovych2007b,Boyko&Patera&Popovych2007c}
the following technical lemma on matrices is used in the proof of the below theorem.

\begin{lemma}\label{LemmaOnEqualitiesWithSubmatrix}
Suppose $1<k<n$.
If $|X^{\varkappa+1,n}_{1,k-1}|\ne0$, then for any $\beta\in\mathbb F$
\begin{gather*}\arraycolsep=0.5ex
\beta-X^{i,i}_{1,k-1}(X^{\varkappa+1,n}_{1,k-1})^{-1}X^{\varkappa+1,n}_{j,j}=
\frac{(-1)^{k+1}}{|X^{\varkappa+1,n}_{1,k-1}|}
\left|\begin{array}{lc} X^{i,i}_{1,k-1} & \beta \\[1ex]
X^{\varkappa+1,n}_{1,k-1}& X^{\varkappa+1,n}_{j,j} \end{array}\!\right|.
\end{gather*}
In particular,
$x_{\varkappa k}-X^{\varkappa,\varkappa}_{1,k-1}(X^{\varkappa+1,n}_{1,k-1})^{-1}X^{\varkappa+1,n}_{k,k}=
(-1)^{k+1} |X^{\varkappa+1,n}_{1,k-1}|^{-1} |X^{\varkappa,n}_{1,k}|$.
Analogously
\begin{gather*}
\left(x_{\varkappa j}-X^{\varkappa,\varkappa}_{1,k-1}(X^{\varkappa+1,n}_{1,k-1})^{-1}X^{\varkappa+1,n}_{j,j}\right)
\left(x_{jk}-X^{j,j}_{1,k-1}(X^{\varkappa+1,n}_{1,k-1})^{-1}X^{\varkappa+1,n}_{k,k}\right)
\\[1ex]\arraycolsep=.5ex
\qquad=\frac{1}{|X^{\varkappa+1,n}_{1,k-1}|}
\left|\begin{array}{lc} X^{j,j}_{1,k} & \beta \\[1ex] X^{\varkappa,n}_{1,k}& X^{\varkappa,n}_{j,j} \end{array}\!\right|+
\frac{|X^{\varkappa,n}_{1,k}|}{|X^{\varkappa+1,n}_{1,k-1}|^2}
\left|\begin{array}{lc} X^{j,j}_{1,k-1} & \beta \\[1ex] X^{\varkappa+1,n}_{1,k-1}& X^{\varkappa+1,n}_{j,j} \end{array}\!\right|.
\end{gather*}
\end{lemma}

\begin{theorem}\label{TheoremOnBasisOfInvsOfCoadjRepresentationOfDiagSolvAlgsWithTriangularNilradical}
Let the parameter matrix~$\gamma$ be of a reduced form.
A basis of $\Inv({\rm Ad}^*_{{\rm T}_\gamma(n)})$ is formed by the expressions%
\footnote{\label{FootnoteOnDomainOfInvsAnsPowers}
These expressions are not defined on the whole space~$\mathfrak t_\gamma^*(n)$.
In particular, the expressions in the second row are well defined only if $|X^{\varkappa,n}_{1,k}|\ne0$, $k=1,\dots,[n/2]$. 
In the complex case with non-integer exponents, a branch of the ln should be fixed and then used for expressing, 
via the exponential function, all powers involved in the expressions in the first row. 
In the real case, these powers are defined for any values of their exponents only for~$x$'s, 
where the determinants being their bases are positive. 
In the general situation of the real case, when an exponent is not an integer or a rational number with odd denominator, 
the corresponding determinant should be replaced by its absolute value.
}
\begin{gather*}
|X^{\varkappa,n}_{1,k}|\prod_{q=1}^{s'\!}|X^{\varkappa_q,n}_{1,k_q}|^{\beta_{qk}}, \quad
k\in\{1, \dots, [n/2]\}\setminus\{k_1 , \dots, k_{s'\!}\},\quad
\\ \arraycolsep=.5ex
x_{p0}+\sum_{k=1}^{\left[\frac n2\right]} \frac{(-1)^{k+1}}{|X^{\varkappa,n}_{1,k}|} (\gamma_{pk}-\gamma_{p,k+1}) \sum_{k<i<\varkappa}
\left|\begin{array}{lc} X^{i,i}_{1,k} & 0 \\[1ex] X^{\varkappa,n}_{1,k}& X^{\varkappa,n}_{i,i} \end{array}\!\right|,
\quad p=s'+1, \ldots, s,
\end{gather*}
where $\beta_{qk}=-\Delta_{qk}/\Delta$,%
\footnote{
In view of their definitions, the parameters $\alpha_{qk}$ and, therefore, $\beta_{qk}$ vanish for any~$q$ and $k<k_1$.  
}
$\Delta=\det(\alpha_{q'\!k_{q''}})_{q'\!,q''\!=1,\dots,s'}=(-1)^{s'}$,
$\Delta_{qk}$ is the determinant obtained from $\Delta$ with change of the column $(\alpha_{q'\!k_q})_{q'\!=1,\dots,s'}$ by
the column $(\alpha_{q'\!k})_{q'\!=1,\dots,s'}$,
\[
\alpha_{qk}:=-\sum_{k'\!=1}^k (\gamma_{q\varkappa'\!}-\gamma_{qk'\!})=-\sum_{k'\!=k_q}^k (\gamma_{q\varkappa'\!}-\gamma_{qk'\!}).
\]
\end{theorem}

\begin{proof}
Under normalization we impose the following constraints on the lifted invariants $\mathcal I_{ij}$, $j<i$:
\[
\mathcal I_{ij}=0 \quad\mbox{if}\quad j<i,\ (i,j)\not=(n-j'+1,j'),\ j'=1,\ldots,\left[\frac{n}2\right].
\]
It means that we do not fix only values of the elements of the lifted invariant matrix~$\mathcal I$,
which are situated on the secondary diagonal under the main diagonal.
The other significant elements of~$\mathcal I$ are put equal to 0.
The choice of just such normalization conditions is a result of a wide preliminary analysis.
It can be justified, in particular, by the structure of the entire automorphism group of~$\mathfrak t_0(n)$,
adduced, e.g., in~\cite{Cao&Tan2003}.

The decision on what to do with the singular lifted invariants~$\mathcal I_{p0}$'s and the
secondary diagonal lifted invariants $\mathcal I_{\varkappa k}$,  $k=1,\dots,[n/2]$, is left for the later discussion,
since it will turn out that necessity of imposing normalization conditions on them depends on values of~$\gamma$.
As shown below, the final normalization in all the cases provides satisfying the conditions of
Proposition~1 and, therefore, is correct.

In view of (triangular) structure of the matrices $B$ and $X$
the formula $\mathcal I=BXB^{-1}$ determining the matrix part of lifted invariants implies that $BX=\mathcal IB$. 
This matrix equality is also significant only for the matrix entries of its left- and right-hand sides~$BX$ and $\mathcal IB$
that underlie the respective main diagonals, i.e., we have the system
\[
e^{\gamma_{pi}\varepsilon_p}x_{ij}+\sum_{i<i'}b_{ii'}x_{i'\!j}
=\mathcal I_{ij}e^{\gamma_{pj}\varepsilon_p}+\sum_{j'<j}\mathcal I_{ij'}b_{j'\!j}, \quad j<i.
\]
For convenience we divide this system under the chosen normalization conditions into four sets of subsystems
\begin{gather*}
S_1^k\colon\qquad e^{\gamma_{p\varkappa}\varepsilon_p}x_{\varkappa j}+\sum_{i'>\varkappa}b_{\varkappa i'}x_{i'\!j}=0, \qquad
i=\varkappa,\quad j<k,\quad k=2,\ldots,\left[\frac{n+1}2\right],
\\[.5ex]
S_2^k\colon\qquad
e^{\gamma_{p\varkappa}\varepsilon_p}x_{\varkappa k}+\sum_{i'>\varkappa}b_{\varkappa i'}x_{i'\!k}
=\mathcal I_{\varkappa k}e^{\gamma_{pk}\varepsilon_p}, \qquad
i=\varkappa,\quad j=k,\quad k=1,\ldots,\left[\frac n2\right],
\\[.5ex]
S_3^k\colon\qquad e^{\gamma_{p\varkappa}\varepsilon_p}x_{\varkappa j}+\sum_{i'>\varkappa}b_{\varkappa i'}x_{i'\!j}
=\mathcal I_{\varkappa k}b_{kj}, \qquad
i=\varkappa,\quad k<j<\varkappa,\quad k=1,\ldots,\left[\frac n2\right]-1,
\\[.5ex]
S_4^k\colon\qquad e^{\gamma_{pk}\varepsilon_p}x_{kj}+\sum_{i'>k}b_{ki'}x_{i'\!j}=0, \qquad
i=k,\quad j<k,\quad k=2,\ldots,\left[\frac n2\right],
\end{gather*}
and solve them one after another.
The subsystem~$S_2^1$ consists of the single equation
\[
\mathcal I_{n1}=x_{n1}e^{(\gamma_{pn}-\gamma_{p1})\varepsilon_p}.
\]
For any fixed $k\in\{2,\dots,[n/2]\}$ the subsystem $S_1^k \cup S_2^k$ is a well-defined system of linear equations with respect to
$b_{\varkappa i'}$, $i'>\varkappa$, and $\mathcal I_{\varkappa k}$.
Analogously, the subsystem  $S_1^k$ for $k=\varkappa=[(n+1)/2]$ in the case of odd~$n$
is a well-defined system of linear equations with respect to $b_{k i'}$, $i'>k$.
The solutions of the above subsystems are expressions of $x_{i'\!j}$, $i'\geqslant\varkappa$, $j<k$, and $\varepsilon_p$:
\begin{gather*}
\mathcal I_{\varkappa k}=(-1)^{k+1}
\frac{|X^{\varkappa,n}_{1,k}|}{|X^{\varkappa+1,n}_{1,k-1}|}\,e^{(\gamma_{p\varkappa}-\gamma_{pk})\varepsilon_p},
\quad k=2, \ldots, \left[\frac n2\right],
\\
B^{\varkappa,\varkappa}_{\varkappa+1,n}=-e^{\gamma_{p\varkappa}\varepsilon_p}X^{\varkappa,\varkappa}_{1,k-1}(X^{\varkappa+1,n}_{1,k-1})^{-1},
\quad k=2, \ldots, \left[\frac {n+1}2\right].
\end{gather*}

After substituting the expressions of $\mathcal I_{\varkappa k}$ and $b_{\varkappa i'}$, $i'>\varkappa$, via $\varepsilon_p$ and $x$'s
into $S_3^k$, we trivially solve the obtained system with respect to $b_{kj}$ as uncoupled system of linear equations:
\begin{gather*}
b_{1j}=e^{\gamma_{p1}\varepsilon_p}\frac{x_{nj}}{x_{n1}}, \quad
1<j<n,
\\[.5ex]
b_{kj}=(-1)^{k+1}e^{\gamma_{pk}\varepsilon_p}
\frac{|X^{\varkappa+1,n}_{1,k-1}|}{|X^{\varkappa,n}_{1,k}|}
\left(x_{\varkappa j}-X^{\varkappa,\varkappa}_{1,k-1}(X^{\varkappa+1,n}_{1,k-1})^{-1}X^{\varkappa+1,n}_{j,j}\right)
=\frac{e^{\gamma_{pk}\varepsilon_p}}{|X^{\varkappa,n}_{1,k}|}
\arraycolsep=0.5ex
\left|\begin{array}{ll} X^{\varkappa,\varkappa}_{1,k-1} & x_{\varkappa j} \\[1ex]
X^{\varkappa+1,n}_{1,k-1}& X^{\varkappa+1,n}_{j,j} \end{array}\!\right|,
\\[.5ex]
k<j<\varkappa,\quad k=2,\ldots,\left[\frac n2\right]-1.
\end{gather*}

Performing the subsequent substitution of the calculated expressions for $b_{kj}$ into $S_4^k$, for any fixed appropriate~$k$
we obtain a well-defined system of linear equations, e.g., with respect to $b_{ki'}$, $i'>\varkappa$.
Its solution is expressed via $x$'s, $b_{k\varkappa}$ and $\varepsilon_p$:
\begin{gather*}
B^{k,k}_{\varkappa+1,n}=-\biggl(e^{\gamma_{pk}\varepsilon_p}X^{k,k}_{1,k-1}+
\sum_{k< j\leqslant\varkappa}b_{kj}X^{j,j}_{1,k-1}\biggr)(X^{\varkappa+1,n}_{1,k-1})^{-1}
\\\phantom{B^{k,k}_{\varkappa+1,n}}
=-b_{k\varkappa}X^{\varkappa,\varkappa}_{1,k-1}(X^{\varkappa+1,n}_{1,k-1})^{-1}
-\frac{e^{\gamma_{pk}\varepsilon_p}}{|X^{\varkappa,n}_{1,k}|}
\sum_{k\leqslant j<\varkappa}\arraycolsep=0.5ex
\left|\begin{array}{ll} X^{\varkappa,\varkappa}_{1,k-1} & x_{\varkappa j} \\[1ex]
X^{\varkappa+1,n}_{1,k-1}& X^{\varkappa+1,n}_{j,j} \end{array}\!\right|
X^{j,j}_{1,k-1}(X^{\varkappa+1,n}_{1,k-1})^{-1},
\\
k=2,\ldots,\left[\frac n2\right].
\end{gather*}

We rewrite the expressions of the lifted invariants~$\mathcal I_{p0}$'s, taking into account the already imposed normalization constraints
(note that $\varkappa=[(n+1)/2]+1$ if $k=[n/2]$):
\begin{gather*}
\mathcal I_{p0}=x_{p0}
+\sum_l\gamma_{pl}\widehat b_{ll}\sum_{l<i}b_{li}x_{il}
+\sum_{k=2}^{\left[\frac{n+1}2\right]}\sum_{j<k}\gamma_{pk}\widehat b_{jk}\sum_{i\geqslant k}b_{ki}x_{ij}
\\\phantom{\mathcal I_{p0}=}
+\sum_{k=1}^{\left[\frac n2\right]}\Biggl(\,\sum_{j<k}+\sum_{k\leqslant j<\varkappa}\,\Biggr)
\gamma_{p\varkappa}\widehat b_{j\varkappa}\sum_{i\geqslant \varkappa}b_{\varkappa i}x_{ij}
\\\phantom{\mathcal I_{p0}}
=x_{p0}
+\sum_l\gamma_{pl}\widehat b_{ll}\sum_{l<i}b_{li}x_{il}
+\sum_{k=1}^{\left[\frac n2\right]}\gamma_{p\varkappa}\mathcal I_{\varkappa k}\sum_{k\leqslant j<\varkappa}b_{kj}\widehat b_{j\varkappa}
\\\phantom{\mathcal I_{p0}}
=x_{p0}
+\sum_{k=1}^{\left[\frac n2\right]}\gamma_{pk}\widehat b_{kk}\Biggl(\,\sum_{k<i\leqslant\varkappa}+\sum_{i>\varkappa}\,\Biggr)b_{ki}x_{ik}
+\sum_{k=1}^{\left[\frac{n+1}2\right]}\gamma_{p\varkappa}\widehat b_{\varkappa\varkappa}\sum_{i>\varkappa}b_{\varkappa i}x_{i\varkappa}
-\sum_{k=1}^{\left[\frac n2\right]}\gamma_{p\varkappa}\widehat b_{\varkappa\varkappa}\mathcal I_{\varkappa k}b_{k\varkappa}.
\end{gather*}
Then we substitute the found expressions for $b$'s and $I_{\varkappa k}$ into the derived expressions of~$\mathcal I_{p0}$'s:
\begin{gather*}
\mathcal I_{p0}=x_{p0}
+\gamma_{p1}e^{-\gamma_{p1}\varepsilon_p}\!\!\sum_{1<i\leqslant n}b_{1i}x_{i1}
+\sum_{k=2}^{\left[\frac n2\right]}\gamma_{pk}e^{-\gamma_{pk}\varepsilon_p}\!\! \sum_{k<i\leqslant\varkappa}b_{ki}
\left(x_{ik}-X^{i,i}_{1,k-1}(X^{\varkappa+1,n}_{1,k-1})^{-1}X^{\varkappa+1,n}_{k,k}\right)
\\\phantom{\mathcal I_{p0}=}
-\sum_{k=2}^{\left[\frac n2\right]}\gamma_{pk}X^{k,k}_{1,k-1}(X^{\varkappa+1,n}_{1,k-1})^{-1}X^{\varkappa+1,n}_{k,k}
+\sum_{k=1}^{\left[\frac{n+1}2\right]}\gamma_{p\varkappa}\widehat b_{\varkappa\varkappa}\sum_{i>\varkappa}b_{\varkappa i}x_{i\varkappa}
-\sum_{k=1}^{\left[\frac n2\right]}\gamma_{p\varkappa}\widehat b_{\varkappa\varkappa}\mathcal I_{\varkappa k}b_{k\varkappa}
\\\phantom{\mathcal I_{p0}}
=x_{p0}
+(\gamma_{p1}-\gamma_{pn})e^{-\gamma_{p1}\varepsilon_p}b_{1n}x_{n1}
+\sum_{k=2}^{\left[\frac n2\right]}(\gamma_{pk}-\gamma_{p\varkappa})e^{-\gamma_{pk}\varepsilon_p}b_{k\varkappa}(-1)^{k+1}
\frac{|X^{\varkappa,n}_{1,k}|}{|X^{\varkappa+1,n}_{1,k-1}|}
\\\phantom{\mathcal I_{p0}=}
-\sum_{k=2}^{\left[\frac n2\right]}
\gamma_{pk}X^{k,k}_{1,k-1}(X^{\varkappa+1,n}_{1,k-1})^{-1}X^{\varkappa+1,n}_{k,k}
-\sum_{k=2}^{\left[\frac{n+1}2\right]}
\gamma_{p\varkappa} X^{\varkappa,\varkappa}_{1,k-1}(X^{\varkappa+1,n}_{1,k-1})^{-1}X^{\varkappa+1,n}_{\varkappa,\varkappa}
\\\phantom{\mathcal I_{p0}=}\arraycolsep=.5ex
+\sum_{k=1}^{\left[\frac n2\right]} \frac{(-1)^{k+1}\gamma_{pk}}{|X^{\varkappa,n}_{1,k}|}\sum_{k<i<\varkappa}
\left|\begin{array}{lc} X^{i,i}_{1,k} & 0 \\[1ex] X^{\varkappa,n}_{1,k}& X^{\varkappa,n}_{i,i} \end{array}\!\right|
+\sum_{k=2}^{\left[\frac n2\right]} \frac{(-1)^{k+1}\gamma_{pk}}{|X^{\varkappa+1,n}_{1,k-1}|}\sum_{k<i<\varkappa}
\left|\begin{array}{lc} X^{i,i}_{1,k-1} & 0 \\[1ex] X^{\varkappa+1,n}_{1,k-1}& X^{\varkappa+1,n}_{i,i} \end{array}\!\right|.
\end{gather*}

Below it is essential for consideration that $\gamma$ is of a reduced form.
For any fixed $q\in\{1,\dots,s'\}$ the lifted invariant $\mathcal I_{q0}$ necessarily depends on
the parameter $b_{k_q\varkappa_q}$ which are not, under already possessed normalization conditions,
in the expressions of the other lifted invariants.
Hence in this case we should use additional normalization conditions constraining $\mathcal I_{q0}$,
e.g., $\mathcal I_{q0}=0$.
It gives an expression for $b_{k_q\varkappa_q}$, $q=1,\dots,s'$, via $x$'s, other $b_{k\varkappa}$'s and $\varepsilon_p$.
The exact form of the expression for $b_{k_q\varkappa_q}$ is inessential.
Since $\gamma_{pk}=\gamma_{p\varkappa}$ for $p>s'$, 
the expressions for $\mathcal I_{p0}$ with  $p>s'$ depend on no group parameters and, therefore, are invariants.
Let us show that the above formula for $\mathcal I_{p0}$ with $p>s'$ gives the second subset of invariants from the statement of the theorem.
We take into account the supposition on $\gamma$ and permute terms in this formula:
\begin{gather*}\arraycolsep=.5ex
\mathcal I_{p0}=x_{p0}
+\sum_{k=1}^{\left[\frac n2\right]} \frac{(-1)^{k+1}\gamma_{pk}}{|X^{\varkappa,n}_{1,k}|}\sum_{k<i<\varkappa}
\left|\begin{array}{lc} X^{i,i}_{1,k} & 0 \\[1ex] X^{\varkappa,n}_{1,k}& X^{\varkappa,n}_{i,i} \end{array}\!\right|
+\sum_{k=2}^{\left[\frac n2\right]} \frac{(-1)^{k+1}\gamma_{pk}}{|X^{\varkappa+1,n}_{1,k-1}|}\sum_{k<i<\varkappa}
\left|\begin{array}{lc} X^{i,i}_{1,k-1} & 0 \\[1ex] X^{\varkappa+1,n}_{1,k-1}& X^{\varkappa+1,n}_{i,i} \end{array}\!\right|
\\\phantom{\mathcal I_{p0}=}
-\sum_{k=2}^{\left[\frac n2\right]}
\gamma_{pk}X^{k,k}_{1,k-1}(X^{\varkappa+1,n}_{1,k-1})^{-1}X^{\varkappa+1,n}_{k,k}
-\left(\sum_{k=2}^{\left[\frac n2\right]}+\sum_{k=\left[\frac n2\right]+1}^{\left[\frac{n+1}2\right]}\right)
\gamma_{pk} X^{\varkappa,\varkappa}_{1,k-1}(X^{\varkappa+1,n}_{1,k-1})^{-1}X^{\varkappa+1,n}_{\varkappa,\varkappa}.
\end{gather*}
For convenience, denote the summation complexes in the derived formula by $\Sigma_1$, \dots, $\Sigma_5$ 
(two and three complexes in the first and second formula's rows, respectively). 
The complex~$\Sigma_5$ contains no summands (resp.\ one summand) if $n$ is even (resp.\ odd).
Applying the first part of Lemma~\ref{LemmaOnEqualitiesWithSubmatrix} for $\beta=0$, we reduce summands of $\Sigma_3$, $\Sigma_4$ and $\Sigma_5$ 
to the form similar to that of summands of $\Sigma_2$. 
We attach the modified summands to $\Sigma_2$ and thus extend the summation intervals to $k,\dots,\varkappa$ for~$i$ (using summands of~$\Sigma_3$ and~$\Sigma_4$)
and to $2,\dots,[n/2]+1$ for~$k$ (using the summand of~$\Sigma_5$ if $n$ is odd; 
the extension is not needed if $n$ is even), 
\begin{gather*}\arraycolsep=.5ex
\mathcal I_{p0}=x_{p0}
+\sum_{k=1}^{\left[\frac n2\right]} \frac{(-1)^{k+1}\gamma_{pk}}{|X^{\varkappa,n}_{1,k}|}\sum_{k<i<\varkappa}
\left|\begin{array}{lc} X^{i,i}_{1,k} & 0 \\[1ex] X^{\varkappa,n}_{1,k}& X^{\varkappa,n}_{i,i} \end{array}\!\right|
+\sum_{k=2}^{\left[\frac n2\right]+1} \frac{(-1)^{k+1}\gamma_{pk}}{|X^{\varkappa+1,n}_{1,k-1}|}
\sum_{k\leqslant i\leqslant\varkappa}
\left|\begin{array}{lc} X^{i,i}_{1,k-1} & 0 \\[1ex] X^{\varkappa+1,n}_{1,k-1}& X^{\varkappa+1,n}_{i,i} \end{array}\!\right|.
\end{gather*}
The shifting of the index~$k$ by $-1$ in the last sum, $k'=k-1$ and thus $\varkappa'=\varkappa+1$, changes 
the summation intervals to $1,\dots,[n/2]$ for~$k'$ and to 
$k'+1,\dots,\varkappa'-1$ for~$i$. 
The recombination of terms leads to the required expression. 

Then, we take $\hat{\mathcal I}_1=\mathcal I_{n1}$
and the combinations $\hat{\mathcal I}_k=(-1)^{k+1}\mathcal I_{\varkappa k}\hat{\mathcal I}_{k-1}$,
$k=2,\dots,[n/2]$, i.e.,
\[
\hat{\mathcal I}_k=|X^{\varkappa,n}_{1,k}|e^{-\alpha_{qk}\varepsilon_q},\qquad
\alpha_{qk}:=-\sum_{k'\!=1}^k (\gamma_{q\varkappa'\!}-\gamma_{qk'\!})=-\sum_{k'\!=k_q}^k (\gamma_{q\varkappa'\!}-\gamma_{qk'\!}),
\qquad k=1,\dots,[n/2].
\]
Since $\hat{\mathcal I}_{k_q}$ depends only on $\varepsilon_q$, \dots, $\varepsilon_{s'}$ among~$\varepsilon$'s and
$\partial\hat{\mathcal I}_{k_q}/\partial\varepsilon_q=-1$ for any fixed~$q$,
the Jacobian \smash{$|\partial\hat{\mathcal I}_{k_q}/\partial\varepsilon_{q'\!}|$} does not vanish, 
\smash{$|\partial\hat{\mathcal I}_{k_q}/\partial\varepsilon_{q'\!}|=(-1)^{s'}$},
and thus we should impose $s'$ more normalization conditions 
\smash{$\hat{\mathcal I}_{k_q}=1$} or \smash{$\hat{\mathcal I}_{k_q}=\mathop{\rm sgn}|X^{\varkappa_q,n}_{1,k_q}|$} 
in the complex or real case, respectively; cf.\ footnote~\ref{FootnoteOnDomainOfInvsAnsPowers}.
After solving them with respect to $\varepsilon_q$ and substituting the obtained expressions into the other \smash{$\hat{\mathcal I}_k$}'s,
we obtain the first subset of invariants from the statement of the theorem.

Under the normalization we express the non-normalized lifted invariants via only $x$'s and
compute a part of the parameters $b$'s and $\varepsilon$'s of the coadjoint action
via $x$'s and the other $b$'s and $\varepsilon$'s.
The expressions in the obtained tuples of invariants are functionally independent.
No equations involving only $x$'s are obtained.
In view of Proposition~1, this implies
that the choice of normalization constraints, which depends on values of $\gamma$, is correct.
That is why the number of the found functionally independent invariants is maximal, i.e.,
they form bases of $\Inv({\rm Ad}^*_{{\rm T}_\gamma(n)})$.
\end{proof}

\begin{corollary}\label{CorollaryOnRelatedInvsOfDiagSolvAlgsWithTriangularNilradical}
$|X^{\varkappa,n}_{1,k}|$, $k=1,\dots,[n/2]$, are functionally independent relative invariants
of ${\rm Ad}^*_{{\rm T}_\gamma(n)}$ for any admissible value of~$\gamma$.
\end{corollary}

See, e.g., \cite{Olver1995} for the definition of relative invariants.

\section{Algebra invariants}\label{SectionInvariants}

Let us reformulate Theorem~\ref{TheoremOnBasisOfInvsOfCoadjRepresentationOfDiagSolvAlgsWithTriangularNilradical}
in terms of generalized Casimir operators.

\begin{theorem}\label{TheoremOnBasisOfInvsOfDiagSolvAlgsWithTriangularNilradical}
Let the parameter matrix~$\gamma$ be of a reduced form.
A basis of~$\Inv(\mathfrak t_\gamma(n))$ is formed by the expressions
\begin{gather*}
|\mathcal E^{1,k}_{\varkappa,n}|\prod_{q=1}^{s'\!}|\mathcal E^{1,k_q}_{\varkappa_q,n}|^{\beta_{qk}}, \quad
k\in\{1, \dots, [n/2]\}\setminus\{k_1 , \dots, k_{s'\!}\},
\\ \arraycolsep=.5ex
f_p+\sum_{k=1}^{\left[\frac n2\right]} \frac{(-1)^{k+1}}{|\mathcal E^{1,k}_{\varkappa,n}|}
(\gamma_{pk}-\gamma_{p,k+1}) \sum_{k<i<\varkappa}
\left|\begin{array}{lc} \mathcal E^{1,k}_{i,i} & \mathcal E^{1,k}_{\varkappa,n} \\[1ex]
0 & \mathcal E^{i,i}_{\varkappa,n} \end{array}\!\right|,
\quad p=s'+1, \ldots, s,
\end{gather*}
where 
$\varkappa:=n-k+1$, $\varkappa_q:=n-k_q+1$;
$\mathcal E^{i_1,i_2}_{j_1,j_2}$, $i_1\leqslant i_2$, $j_1\leqslant j_2$, denotes the matrix $(e_{ij})^{i=i_1,\ldots,i_2}_{j=j_1,\ldots,j_2}$;
$\beta_{qk}=-\Delta_{qk}/\Delta$, $\Delta=\det(\alpha_{q'\!k_{q''}})_{q'\!,q''\!=1,\dots,s'}=(-1)^{s'}$,
$\Delta_{qk}$ is the determinant obtained from $\Delta$
with change of the column $(\alpha_{q'\!k_q})_{q'\!=1,\dots,s'}$ by
the column $(\alpha_{q'\!k})_{q'\!=1,\dots,s'}$,
\[
\alpha_{qk}:=-\sum_{k'\!=1}^k (\gamma_{q,n-k'+1}-\gamma_{qk'\!})=-\sum_{k'\!=k_q}^k (\gamma_{q,n-k'+1}-\gamma_{qk'\!}).
\]
\end{theorem}

\begin{proof}
Expanding the determinants in each element of the first tuple of invariants from
Theorem~\ref{TheoremOnBasisOfInvsOfCoadjRepresentationOfDiagSolvAlgsWithTriangularNilradical},
we obtain an expression of $x$'s containing only such coordinate functions that
the corresponding basis elements commute each to other. 
Therefore, the symmetrization procedure is trivial.
Since $x_{ij}\sim e_{ji}$, $j<i$, hereafter it is necessary to transpose the matrices
in the obtained expressions of invariants for representation improvement. 
Finally we construct the first part of the basis of~$\Inv(\mathfrak t_\gamma(n))$ from the statement. 

The symmetrization procedure for the second tuple of invariants presented in
Theorem~\ref{TheoremOnBasisOfInvsOfCoadjRepresentationOfDiagSolvAlgsWithTriangularNilradical} 
also can be assumed trivial. To show this, we again expand all the determinants.  
Only the monomials of the determinants 
\[\arraycolsep=.5ex
\left|\begin{array}{lc} X^{i,i}_{1,k} & 0 \\[1ex] X^{\varkappa,n}_{1,k}& X^{\varkappa,n}_{i,i} \end{array}\!\right|, 
\quad k\in\{1, \dots, [n/2]\},\quad i=k,\dots,\varkappa, 
\]
contain coordinate functions associated with noncommuting basis elements of the algebra $\mathfrak t_\gamma(n)$.
More precisely, each of the monomials includes two such coordinate functions, namely, 
$x_{ii'\!}$ and $x_{j'\!i}$ for some values $i'\in\{1,\dots,k\}$ and $j'\in\{\varkappa,\dots,n\}$. 
It is sufficient to make only the symmetrization of the corresponding pairs of basis elements. 
As a result, after the symmetrization and the transposition of the matrices we obtain the following expressions 
for the invariants of $\mathfrak t_\gamma(n)$ corresponding to the invariants of the second tuple from
Theorem~\ref{TheoremOnBasisOfInvsOfCoadjRepresentationOfDiagSolvAlgsWithTriangularNilradical}:
\[
f_p+\sum_{k=1}^{\left[\frac n2\right]} \frac{(-1)^{k+1}}{|\mathcal E^{1,k}_{\varkappa,n}|}
(\gamma_{pk}-\gamma_{p,k+1}) \sum_{k<i<\varkappa}\sum_{i'=1}^k\sum_{j'=\varkappa}^n
\frac{e_{i'\!i}e_{ij'}+e_{ij'}e_{i'\!i}}2(-1)^{i'\!j'}\bigl|\mathcal E^{1,k;\hat i'}_{\varkappa,n;\hat j'}\bigr|,
\]
where $p=s'+1,\ldots,s$ and $\bigl|\mathcal E^{1,k;\hat i'}_{\varkappa,n;\hat j'}\bigr|$ denotes 
the minor of the matrix $\mathcal E^{1,k}_{\varkappa,n}$ complementary to the element $e_{i'\!j'}$. 
Since $e_{i'\!i}e_{ij'}=e_{ij'}e_{i'\!i}+e_{i'\!j'}$, then 
\[
\sum_{i'=1}^k\sum_{j'=\varkappa}^n
\frac{e_{i'\!i}e_{ij'}+e_{ij'}e_{i'\!i}}2(-1)^{i'\!j'}\bigl|\mathcal E^{1,k;\hat i'}_{\varkappa,n;\hat j'}\bigr|=
\arraycolsep=.5ex
\left|\begin{array}{lc} \mathcal E^{1,k}_{i,i} & \mathcal E^{1,k}_{\varkappa,n} \\[1ex]
0 & \mathcal E^{i,i}_{\varkappa,n} \end{array}\!\right|
\pm\frac12|\mathcal E^{1,k}_{\varkappa,n}|, 
\]
where we have to take the sign `$+$' (resp. `$-$') if 
the elements of~$\mathcal E^{1,k}_{i,i}$ are placed after (resp. before) 
the elements of~$\smash{\mathcal E^{i,i}_{\varkappa,n}}$ in all the relevant monomials. 
Up to constant summands, 
this results in the expressions for the elements of the second part of the invariant basis adduced in the statement. 
These expressions are formally derived from the corresponding expressions from 
Theorem~\ref{TheoremOnBasisOfInvsOfCoadjRepresentationOfDiagSolvAlgsWithTriangularNilradical}
by the replacement $x_{ij}\to e_{ji}$ and $x_{p0}\to f_p$ and the transposition of all matrices. 
That is why we assume that the symmetrization procedure is trivial in the sense described.
Let us emphasize that 
a uniform order of elements from $\mathcal E^{1,k}_{i,i}$ and $\mathcal E^{i,i}_{\varkappa,n}$ 
has to be fixed in all the monomials under usage of the `non-symmetrized' form of invariants. 
\end{proof}

For a matrix~$\gamma$ in a reduced form 
we denote $K:=\{1, \dots, [n/2]\}\setminus\{k_1 , \dots, k_{s'\!}\}$;
see the notation in Proposition~\ref{PropositionOnReducedFormOfParameterMatrix}.

\begin{corollary}\label{CorollaryOnRationalInvsOfDiagSolvAlgsWithTriangularNilradical}
The algebra~$\mathfrak t_\gamma(n)$ with~$\gamma$ in a reduced form admits a rational basis of invariants 
if and only if
$\beta_{qk}\in\mathbb Q$ for all $k\in K$
and all $q\in\{1, \dots, s'\}$. 
\end{corollary}

\begin{corollary}\label{CorollaryOnPolynomialInvsOfDiagSolvAlgsWithTriangularNilradical}
The algebra~$\mathfrak t_\gamma(n)$  with~$\gamma$ in a reduced form admits a polynomial basis of invariants 
(i.e., a basis consisting of Casimir operators)
if and only if
$\beta_{qk}\in\mathbb Q$ for all $k\in K$, 
$\gamma_{pk_q}=\gamma_{p,k_q+1}$ for $p=s'+1, \ldots, s$ and each~$q$ with $\beta_{qk}=0$, $k\in K$, 
and there exist positive $\lambda_k\in\mathbb Q$, $k\in K$, such that 
$\sum_{k\in K}\beta_{qk}\lambda_k>0$ for any other~$q$. 
\end{corollary}

We can reformulate Corollary~\ref{CorollaryOnPolynomialInvsOfDiagSolvAlgsWithTriangularNilradical} 
using known results on compatibility of systems of homogeneous linear inequality in terms of associated matrices \cite{Blumenthal1952,Dines1919}.
We can also derive various simpler particular conditions that are sufficient for the existence of a polynomial basis of~$\Inv(\mathfrak t_\gamma(n))$:
\begin{enumerate}\itemsep=0ex
\item 
For some fixed~$k\in K$, $\beta_{qk}>0$ for all $q$.
\item 
$\beta_{qk}\geqslant0$ for all $k\in K$ and for all~$q$, 
and, if $s'<s$, then for each~$q$ with $\gamma_{pk_q}-\gamma_{p,k_q+1}\ne0$ there exists $k\in K$ such that $\beta_{qk}>0$.
\item 
There exist $l_r\in K$, where $r=1,\dots, s''$ ($s''\leqslant s'$), such that $\beta_{ql_r}\geqslant0$, 
for each~$q$, where $\beta_{qk}<0$ for some~$k\in K$, there exists~$r$ with $\beta_{ql_r}>0$, 
and, if $s'<s$, then for each~$q$ with $\gamma_{pk_q}-\gamma_{p,k_q+1}\ne0$ there exists $k\in K$ such that $\beta_{qk}>0$.
\end{enumerate}  

\begin{remark}\label{NoteOnAlgReformulationOfCondOfExtensionOfInvSetForDiagSolvAlgsWithTriangularNilradical}
It follows from Theorem~\ref{TheoremOnBasisOfInvsOfDiagSolvAlgsWithTriangularNilradical} that
the cardinality $N_{\mathfrak t_\gamma(n)}$ of fundamental invariants
of the algebra $\mathfrak t_\gamma(n)$ equals to $[n/2]+s-2s'$, where
$s$ is the number of nilindependent elements and
\[
s'=\rank(\gamma_{p\varkappa}-\gamma_{pk})^{p=1,\dots,s}_{k=1,\dots,[n/2]}
=\rank(\gamma_{p\varkappa}-\gamma_{pk})^{p=1,\dots,s}_{k=1,\dots,n}.
\]
For any fixed~$s$ the cardinality $N_{\mathfrak t_\gamma(n)}$ is maximal if $s'$ has the minimally possible value.
In the case $s\in\{1,\dots,[n/2]\}$ such value is $s'=0$ and, therefore, $N_{\mathfrak t_\gamma(n)}=[n/2]+s$.
It means that $\gamma_{pk}=\gamma_{p\varkappa}$ for all $k\in\{1,\dots,[n/2]\}$ and all $p\in\{1,\dots,s\}$.
This condition can be reformulated in terms of commutators in the following way.
Any nilindependent element commute with the `nilpotent' basis elements
$e_{k\varkappa}$, $k=1,\dots,[n/2]$, lying on the significant part of the secondary diagonal
of the basis `matrix'~$\mathcal E$, i.e., $[f_p,e_{k\varkappa}]=0$, $k=1,\dots,[n/2]$.
If $s\in\{[n/2]+1,\dots,n-1\}$ the minimal value of $s'$ is $s'=s-[n/2]$ and, therefore,
$N_{\mathfrak t_\gamma(n)}=3[n/2]-s$. It is equivalent to the condition that
$[n/2]$ nilindependent elements of the algebra commute with the basis elements $e_{k\varkappa}$, $k=1,\dots,[n/2]$.
\end{remark}

\begin{remark}
\looseness=-1
The elements lying on the secondary diagonal of the matrix of lifted invariants play
a singular role under the normalization procedure in all investigated algebras with
the nilradicals isomorphic to $\mathfrak t_0(n)$:
$\mathfrak t_0(n)$ itself and $\mathfrak{st}(n)$~\cite{Boyko&Patera&Popovych2007b} as well as
$\mathfrak t_\gamma(n)$ studied in this paper.
(More precisely, in~\cite{Boyko&Patera&Popovych2007b} the normalization procedure was realized
for $\mathfrak t(n)$ and then the results on invariants were extended to $\mathfrak{st}(n)$.)
Reasons of such singularity were not evident from the consideration in~\cite{Boyko&Patera&Popovych2007b}.
Remark~\ref{NoteOnAlgReformulationOfCondOfExtensionOfInvSetForDiagSolvAlgsWithTriangularNilradical}
gives an explanation for it and justifies naturalness of the chosen normalization conditions.
\end{remark}

\section{Particular cases}\label{SectionPartialCases}

Theorem~\ref{TheoremOnBasisOfInvsOfDiagSolvAlgsWithTriangularNilradical} includes,
as particular cases, known results on invariants of
the nilpotent algebra of strictly upper triangular matrices $\mathfrak t_0(n)$
\cite{Boyko&Patera&Popovych2007,Boyko&Patera&Popovych2007b,Tremblay&Winternitz2001},
the solvable algebras $\mathfrak{st}(n)$ and $\mathfrak t(n)$
of special upper and non-strictly upper triangular matrices
\cite{Boyko&Patera&Popovych2007b,Tremblay&Winternitz2001}
and the solvable algebras with the nilradical isomorphic to $\mathfrak t_0(n)$ and one nilindependent element
\cite{Boyko&Patera&Popovych2007c,Tremblay&Winternitz2001}.
We show this below, giving additional comments and rewriting invariants in bases
which are more appropriate for the special cases.

Let us remind that $N_\mathfrak g$ denotes
the maximal number of functionally independent invariants in the set $\Inv({\rm Ad}^*_G)$
of invariants of ${\rm Ad}^*_G$, where $G$ is the connected Lie group associated with the Lie algebra~$\mathfrak g$. 
We use the short `non-symmetrized' form for certain basis invariants, where it is uniformly assumed  
that in all monomials elements of $\mathcal E^{1,k}_{i,i}$ is placed before (or after) 
elements of $\smash{\mathcal E^{i,i}_{\varkappa,n}}$. 
See the proof of Theorem~\ref{TheoremOnBasisOfInvsOfDiagSolvAlgsWithTriangularNilradical} for details.

The algebra $\mathfrak t_0(n)$ has no nilindependent elements, i.e., for it $s=0$ and
$|X^{\varkappa,n}_{1,k}|$, $k=1,\dots,[n/2]$, are functionally independent absolute invariants
of ${\rm Ad}^*_{{\rm T}_0(n)}$.

\begin{corollary}\label{CorollaryOnBasisOfInvsOfAlgOfStrictlyUpperTriangularMatrices}
$N_{\mathfrak t_0(n)}=[n/2]$.
A basis of~$\Inv(\mathfrak t_0(n))$ is formed by the Casimir operators (i.e., polynomial invariants)
\[
\det(e_{ij})^{i=1,\ldots,k}_{j=n-k+1,\ldots,n}, \quad k=1, \ldots, \left[\frac n2\right].
\]
\end{corollary}

In the case of one nilindependent element ($s=1$) we can omit the subscript of~$f$ and the first subscript of~$\gamma$.
There are two different cases depending on the value of $s'$ which can be either~0 or~1.
The statement on invariant can be easily formulated even for the unreduced form of~$\gamma$.

\begin{corollary}\label{CorollaryOnBasisOfInvsOfAlgsWithTriangularNilradicalAnd1NonnilpElement}
Let $s=1$. If additionally $s'=0$, i.e., $\gamma_k=\gamma_\varkappa$ for all $k\in\{1,\dots,[n/2]\}$, then
$N_{\mathfrak t_0(n)}=[n/2]+1$ and
a basis of~$\Inv(\mathfrak t_\gamma(n))$ is formed by the expressions
\[\arraycolsep=.5ex
|\mathcal E^{1,k}_{\varkappa,n}|, \quad k=1, \ldots, \left[\frac n2\right], \qquad
f+\sum_{k=1}^{\left[\frac n2\right]} \frac{(-1)^{k+1}}{|\mathcal E^{1,k}_{\varkappa,n}|} (\gamma_k-\gamma_{k+1}) \sum_{i=k+1}^{n-k}
\left|\begin{array}{lc} \mathcal E^{1,k}_{i,i} & \mathcal E^{1,k}_{\varkappa,n} \\[1ex] 0 & \mathcal E^{i,i}_{\varkappa,n} \end{array}\!\right|.
\]
Hereafter $\varkappa:=n-k+1$,
$\mathcal E^{i_1,i_2}_{j_1,j_2}$, $i_1\leqslant i_2$, $j_1\leqslant j_2$, denotes the matrix $(e_{ij})^{i=i_1,\ldots,i_2}_{j=j_1,\ldots,j_2}$.

Otherwise $s'=1$, $N_{\mathfrak t_0(n)}=[n/2]-1$ and
a basis of~$\Inv(\mathfrak t_\gamma(n))$ consists of the invariants
\begin{gather*}
|\mathcal E^{1,k}_{\varkappa,n}|, \quad k=1, \ldots, k_0-1, \qquad
|\mathcal E^{1,k}_{\varkappa,n}|\,|\mathcal E^{1,k_0}_{\varkappa_0,n}|^{\beta_k},
\quad k=k_0+1,\ldots,\left[\frac n2\right],
\end{gather*}
where $k_0$ the minimal value of $k$ for which $\gamma_{k}\ne\gamma_{\varkappa}$ and
\[
\beta_k=-\sum_{i=k_0}^k\frac{\gamma_{n-i+1}-\gamma_i}{\gamma_{n-k_0+1}-\gamma_{k_0}}.
\]
\end{corollary}

The basis constructed for the first case is formed by $[n/2]$ Casimir operators and a nominally rational invariant.
The latter invariant can be replaced by the product of it and
the Casimir operators $|\mathcal E^{1,k}_{\varkappa,n}|$, $k=1, \ldots, [n/2]$.
This product is more complicated but polynomial.
Therefore, under the conditions $s=1$, $s'=0$ the algebra $\mathfrak t_\gamma(n)$ possesses a polynomial fundamental invariant.

In the second case $\Inv(\mathfrak t_\gamma(n))$ has a rational basis if and only if
$\beta_k\in\mathbb Q$ for all $k\in\{k_0,\dots,[n/2]\}$.
Under this condition the obtained basis consists of $k_0-1$ Casimir operators and $[n/2]-k_0$ rational invariants.
If additionally $\beta_k\geqslant0$ for all $k\in\{k_0,\dots,[n/2]\}$, then the whole basis is polynomial.

Note that for both the cases of~$b$ (i.e., for both $b=-1$ and $b\ne-1$)
the results on the algebra~$\mathfrak{g}_{4.8}^{b}$ adduced in Section~\ref{SectionIllustrativeExample} 
are easily derived from Corollary~\ref{CorollaryOnBasisOfInvsOfAlgsWithTriangularNilradicalAnd1NonnilpElement} 
via fixing $n=3$, then identifying $e_1\sim e_{13}$,  $e_2\sim e_{12}$, $e_3\sim e_{23}$ and $e_4\sim f$ and
putting $\gamma_1=-1$, $\gamma_2=0$ and $\gamma_3=b$. 

In the case of the maximal number $s=n-1$ of nilindependent elements the algebra $\mathfrak t_\gamma(n)$
is isomorphic to the algebra $\mathfrak{st}(n)$ of special upper triangular matrices~\cite{Boyko&Patera&Popovych2007b}.
For the matrix~$\gamma$ associated with this algebra, we have 
$\smash{
s'=\rank(\gamma_{p\varkappa}-\gamma_{pk})^{p=1,\dots,s}_{k=1,\dots,[n/2]}=[n/2].}
$
Therefore, $\mathfrak{st}(n)$ has no invariants depending only on elements of the nilradical.
The number of zero rows in the matrix $(\gamma_{p\varkappa}-\gamma_{pk})^{p=1,\dots,s}_{k=1,\dots,[n/2]}$
after reduction of~$\gamma$ should equal to $s-s'=n-1-[n/2]=[(n-1)/2]$.
We choose the basis in $\mathfrak{st}(n)$, which is formed by the elements of the canonical basis of the nilradical
and nilindependent elements $f_p$, $p=1,\dots,n-1$, corresponding to the matrix~$\gamma$ with
\[
\gamma_{pi}=\frac{n-p}n,\quad i=1,\dots,p,\qquad
\gamma_{pi}=-\frac pn,\quad i=p+1,\dots,n.
\]
The commutation relations of $\mathfrak {st}(n)$ in the chosen basis are
\begin{gather*}
[e_{ij},e_{i'\!j'}]=\delta_{i'\!j}e_{ij'}-\delta_{ij'}e_{i'\!j}, \qquad i<j,\quad i'<j'; \\
[f_k,f_{k'}]=0,  \qquad k,k'=1,\dots, n-1;\\
[f_k,e_{ij}]=0, \qquad i<j\leqslant k \quad\text{or}\quad k\leqslant i<j; \\
[f_k,e_{ij}]=e_{ij}, \qquad i\leqslant k\leqslant j,\quad  i<j.
\end{gather*}
Then we pass to the basis in which the matrix~$\gamma$ is of a reduced form.
We denote the reduced form by~$\gamma'$.
Only the part of the new basis, which
corresponds to the zero rows of $(\gamma'_{p\varkappa}-\gamma'_{pk})^{p=1,\dots,s}_{k=1,\dots,[n/2]}$,
is essential for finding a fundamental invariant of $\mathfrak{st}(n)$.
As this part, we can take the set consisting of the elements $f'_{s'\!+p}=f_p-f_{n-p}$, $p=1,\dots,[(n-1)/2]$.
Indeed, they are linearly independent and
\[
\gamma'_{s'\!+p,i}=-2\frac pn,\quad i=p+1,\dots,n-p,\qquad
\gamma_{s'\!+p,i}=\frac{n-2p}n\quad \mbox{otherwise}.
\]
Note also that under $p=1,\dots,[(n-1)/2]$ and $k=1,\dots,[n/2]$ the expression
$\gamma'_{s'\!+p,k}-\gamma'_{s'\!+p,k+1}$ equals to 1 if $k=p$ and vanishes otherwise.

\begin{corollary}\label{CorollaryOnBasisOfInvsOfAlgsOfSpecialUpperTriangularMatrices}
$N_{\mathfrak{st}(n)}=[(n-1)/2]$.
A basis of $\Inv(\mathfrak{st}(n))$ consists of the rational invariants
\[\arraycolsep=.5ex
\check{\mathcal I}_k=f_k-f_{n-k}+
\frac{(-1)^{k+1}}{|\mathcal E^{1,k}_{\varkappa,n}|}\displaystyle\sum_{j=k+1}^{n-k}
\left|\begin{array}{ll} \mathcal E^{1,k}_{j,j} & \mathcal E^{1,k}_{\varkappa,n} \\[1ex]
0 & \mathcal E^{j,j}_{\varkappa,n} \end{array}\!\right|,
\quad k=1, \ldots, \left[\frac {n-1}2\right],
\]
where $\mathcal E^{i_1,i_2}_{j_1,j_2}$, $i_1\leqslant i_2$, $j_1\leqslant j_2$, 
denotes the matrix $(e_{ij})^{i=i_1,\ldots,i_2}_{j=j_1,\ldots,j_2}$,
and $\varkappa:=n-k+1$.
\end{corollary}

The algebra $\mathfrak t(n)$ of non-strictly upper triangular matrices stands alone from the considered algebras
since the nilradical of $\mathfrak t(n)$ is wider than $\mathfrak t_0(n)$.
Similarly to $\mathfrak t_0(n)$, the algebra $\mathfrak t(n)$ admit the completely matrix interpretations
of a basis and lifted invariants.
Namely, its basis elements are convenient to enumerate with the `non-decreasing' pair of indices
similarly to the canonical basis $\{E^n_{ij},\,i\leqslant j\}$ of the isomorphic matrix algebra.
Thus, the basis elements $e_{ij}\sim E^n_{ij}$, $i\leqslant j$, satisfy the commutation relations
$[e_{ij},e_{i'\!j'}]=\delta_{i'\!j}e_{ij'}-\delta_{ij'}e_{i'\!j}$,
where $\delta_{ij}$ is the Kronecker delta.

The center of $\mathfrak t(n)$ is one-dimensional and coincides with the linear span of the sum $e_{11}+\dots+e_{nn}$
corresponding to the identity matrix~$E^n$.
The elements $e_{ij}$, $i<j$, and $e_{11}+\dots+e_{nn}$ form a basis of the nilradical of  $\mathfrak t(n)$,
which is isomorphic to $\mathfrak t_0(n)\oplus \mathfrak a$.
Here $\mathfrak a$ is the one-dimensional (Abelian) Lie algebra.

Let $e_{ji}^*$, $x_{ji}$ and $y_{ij}$ denote
the basis element and the coordinate function in the dual space $\mathfrak t^*(n)$ and
the coordinate function in~$\mathfrak t(n)$,
which correspond to the basis element~$e_{ij}$, $i\leqslant j$.
We complete the sets of $x_{ji}$ and $y_{ij}$ to the matrices $X$ and $Y$ with zeros.
Hence $X$ is a lower triangular matrix and $Y$ is an upper triangular one.
In the above notations a fundamental lifted invariant of \raisebox{0ex}[0ex][0ex]{${\rm Ad}^*_{{\rm T}(n)}$} is formed by
the elements $\mathcal I_{ij}$, $j\leqslant i$, of the matrix
$\mathcal I=BXB^{-1}$, where $B$ is an arbitrary matrix from ${\rm T}(n)$ (Lemma~2 of~\cite{Boyko&Patera&Popovych2007b}).
See also Note~3 of~\cite{Boyko&Patera&Popovych2007b} for discussion on essential parameters
in this fundamental lifted invariant.
Due to the matrix representation of lifted invariant, a basis of $\Inv({\rm Ad}^*_{{\rm T}(n)})$ can be
constructed by the normalization procedure in a quite easy way.

At the same time, a basis of $\Inv({\rm Ad}^*_{{\rm T}(n)})$ is obtained from the basis of $\Inv({\rm Ad}^*_{{\rm ST}(n)})$
with attaching the central element $e_{11}+\dots+e_{nn}$.
Indeed, the algebra $\mathfrak t(n)$ is a central extension of~$\mathfrak {st}(n)$, i.e.,
$\mathfrak t(n)=\mathfrak {st}(n)\oplus {\rm Z}(\mathfrak t(n))$,
under the natural embedding of  $\mathfrak {st}(n)$ into $\mathfrak t(n)$.
It is well known that if the Lie algebra $\mathfrak g$ is decomposable into
the direct sum of Lie algebras~$\mathfrak g_1$ and~$\mathfrak g_2$,
then the concatenation of bases of~$\Inv(\mathfrak g_1)$ and~$\Inv(\mathfrak g_2)$ is a basis of~$\Inv(\mathfrak g)$.
A basis of $\Inv({\rm Z}(\mathfrak t(n)))$ obviously consists of only one element, e.g., $e_{11}+\dots+e_{nn}$.
Therefore, the basis cardinality of equals to  $\Inv(\mathfrak {t}(n))$
the basis cardinality of $\Inv(\mathfrak {st}(n))$ plus 1, i.e., $[(n+1)/2]$.
We only combine basis elements and rewrite them in terms of the canonical basis of~$\mathfrak t(n)$.
Namely,
\[
\hat{\mathcal I}_0:=e_{11}+\dots+e_{nn}, \qquad
\hat{\mathcal I}_k=(-1)^{k+1}\check{\mathcal I}_k+(-1)^k\frac{n-2k}n\hat{\mathcal I}_0,
\quad k=1, \ldots, \left[\frac {n-1}2\right].
\]

\begin{corollary}\label{CorollaryOnBasisOfInvsOfAlgOfUpperTriangularMatrices}
$N_{\mathfrak t(n)}=[(n+1)/2]$.
A basis of~$\Inv(\mathfrak t(n))$ consists of the rational invariants
\[\arraycolsep=.5ex
\hat{\mathcal I}_k=
\frac{1}{|\mathcal E^{1,k}_{\varkappa,n}|}\displaystyle\sum_{j=k+1}^{n-k}
\left|\begin{array}{ll} \mathcal E^{1,k}_{j,j} & \mathcal E^{1,k}_{\varkappa,n} \\[1ex]
e_{jj} & \mathcal E^{j,j}_{\varkappa,n} \end{array}\!\right|,
\quad k=0, \ldots, \left[\frac {n-1}2\right],
\]
where $\mathcal E^{i_1,i_2}_{j_1,j_2}$, $i_1\leqslant i_2$, $j_1\leqslant j_2$, denotes the matrix $(e_{ij})^{i=i_1,\ldots,i_2}_{j=j_1,\ldots,j_2}$,
$|\mathcal E^{1,0}_{n+1,n}|:=1$, and $\varkappa:=n-k+1$.
\end{corollary}

Note that in~\cite{Boyko&Patera&Popovych2007b} the inverse way was preferred
due to the simple matrix representation of a fundamental lifted invariant
of $\smash{\rm Ad}^*_{{\rm T}(n)}$.
Namely, at first a basis of~$\Inv(\mathfrak t(n))$ was calculated by the normalization procedure
and then it was used for construction of a basis of~$\Inv(\mathfrak{st}(n))$.

\section{Conclusion and discussion}

In this paper we investigate invariants of solvable Lie algebras
with the nilradicals isomorphic to $\mathfrak t_0(n)$ and `diagonal'
nilindependent elements, using our original pure algebraic
approach~\cite{Boyko&Patera&Popovych2006,Boyko&Patera&Popovych2007}
and the special technique developed
in~\cite{Boyko&Patera&Popovych2007b,Boyko&Patera&Popovych2007c} for
triangular algebras within the framework of this approach. All such
algebras are embedded in $\mathfrak{st}(n)$ as ideals. The number
$s$ of nilindependent elements varies from 0 to $n-1$. In~the
frontier cases $s=0$ and $s=n-1$ the algebras are isomorphic to the
universal algebras $\mathfrak t_0(n)$ and $\mathfrak{st}(n)$, respectively.

The two main steps of the algorithm are the construction of a
fundamental lifted invariant of the coadjoint representation of the
corresponding connected Lie group and the exclusion of parameters
from lifted invariants by the normalization procedure. The
realization of both steps for the algebras under consideration are
more difficult than for the particular cases investigated earlier.
Thus, the constructed fundamental lifted invariant has a more
complicated representation. It is divided into two parts which play
different roles under the normalization. The~part corresponding to
the nilradical admits a simple `matrix' representation which is
important for further consideration. The components from the other
part involves also nilindependent elements and algebra parameters.
That is why the choice of the normalization conditions essentially
depends on algebra parameters that leads to the furcation of
calculations and final results. The partition of the fundamental
lifted invariant induces the partition of normalization
conditions and the associated basis of algebra invariants.

The above obstacles are surmounted due to the optimization of the
applied technique, taking into account properties of the algebras
under consideration, in particular, their standard matrix
representations. This technique involves the choice of special
parameterizations of the inner automorphism groups, the
representation of most of the lifted invariants via matrices and the
natural normalization constraints associated with the algebra
structure. The cardinality of the invariant bases is determined in
process of their construction. Moreover, we only partially constrain
lifted invariants in the beginning of the normalization procedure
and only with conditions without the algebra parameters. Both the total
number of necessary constraints and the additional constraints are
specified before completing of the normalization depending on values of
algebra parameters. As a result of the optimization, excluding the
group parameters $b$'s and $\varepsilon$'s is in fact reduced to
solving linear systems of (algebraic) equations.

We plan to continue investigations of the solvable Lie algebras with
the nilradicals isomorphic to $\mathfrak t_0(n)$ in the general case
where nilindependent elements are not necessarily diagonal. All such
algebras were classified in~\cite{Tremblay&Winternitz1998}, and this
classification can be enhanced with adaptation of known
results~\cite{Cao&Tan2003} on automorphisms of $\mathfrak t_0(n)$.
Unfortunately, it is not understandable as of yet whether the
partial matrix representation of lifted invariants and other tricks
from the developed `triangular' technique will be applicable in
these investigations.

Other possibilities on the usage of the algorithm are outlined in
our previous papers
\cite{Boyko&Patera&Popovych2006,Boyko&Patera&Popovych2007,Boyko&Patera&Popovych2007b,Boyko&Patera&Popovych2007c}.
We hope that the presented results are of interest in the theory of
integrable systems and for labeling of representations of Lie
algebras, as well as other applications, since the algorithm
provides a powerful purely algebraic alternative to the usual method
involving differential equations, and certain ad-hoc methods
developed for special classes of Lie algebras.

\section*{Acknowledgments}
The work of J.\,P. was partially supported by the National Science and Enginee\-ring Research Council of Canada, by the MIND Institute of Costa Mesa, Calif., and by MITACS.
The research of R.\,P. was supported by Austrian Science Fund (FWF), Lise Meitner project M923-N13 and project P25064. 
V.\,B. is grateful for the hospitality extended to him at the Centre de Recherches Math\'ematiques, Universit\'e de Montr\'eal.
The authors thank the referee for useful remarks.

\end{document}